\documentclass[fleqn,usenatbib]{mnras}

\usepackage{newtxtext,newtxmath}

\usepackage[T1]{fontenc}

\DeclareRobustCommand{\VAN}[3]{#2}
\let\VANthebibliography\thebibliography
\def\thebibliography{\DeclareRobustCommand{\VAN}[3]{##3}\VANthebibliography}

\usepackage{graphicx}	
\usepackage{amsmath}	

\newcommand{\pynbody}{{\scriptsize PYNBODY}}
\def\sec#1{Sec.~\ref{sec:#1}} 
\def\apdx#1{Appendix~\ref{sec:#1}} 

\def\Fig#1{Fig.~\ref{fig:#1}}
\def\Table#1{Table~\ref{tab:#1}}
\def\Eq#1{Eq.~\ref{eq:#1}}

\newcommand{\kms}{\ifmmode\,{\rm km}\,{\rm s}^{-1}\else km$\,$s$^{-1}$\fi}
\newcommand{\Rd}{\ifmmode\,R_{\rm d}\else $R_{\rm d}$\fi}
\newcommand{\be}{\begin{equation}}
\newcommand{\ee}{\end{equation}}
\newcommand\ltsima{$\; \buildrel < \over \sim \;$}
\newcommand\ltsim{\lower.5ex\hbox{\ltsima}}
\newcommand\gtsima{$\; \buildrel > \over \sim \;$}
\newcommand\gtsim{\lower.5ex\hbox{\gtsima}}

\newcommand{\magss}{\ifmmode {{{{\rm mag}~{\rm arcsec}}^{-2}}}
             \else {{{mag}$~${arcsec}$^{-2}$}}
             \fi}
\newcommand{\ha}{H$\alpha$}

\def \ion#1#2{#1{\footnotesize{#2}}\relax}
\def \HI{\ion{H}{I}}
\def \littleprime{\ifmmode{\scriptscriptstyle \prime }
     \else{\hbox{$\scriptscriptstyle \prime$ }}\fi}
\def \arcsec{\raise .9ex \hbox{\littleprime\hskip-3pt\littleprime}}

\title[Diversity in Galaxies]{The diversity of spiral galaxies explained}

\author[M. Frosst et al.]{
Matthew Frosst$^{1}$\thanks{E-mail: matthew.frosst@research.uwa.edu.au},
Stéphane Courteau$^{1}$,
Nikhil Arora$^{1}$,
Connor Stone$^{1}$,
Andrea V. Macci\`o$^{2,3,4}$, 
\newauthor
 and Marvin Blank$^{2,3,5}$
\\
$^{1}$Queen’s University, Department of Physics, Engineering Physics and Astronomy, Kingston, Ontario, Canada\\
$^{2}$New York University Abu Dhabi, PO Box 129188, Abu Dhabi, United Arab Emirates\\
$^3$Center for Astro, Particle and Planetary Physics (CAP$^3$), New York University Abu Dhabi\\
$^{4}$Max-Planck-Institut für Astronomie, Königstuhl 17, D-69117 Heidelberg, Germany\\
$^{5}$Institut f\"{u}r Theoretische Physik und Astrophysik, Christian-Albrechts-Universit\"{a}t zu Kiel, Leibnizstr. 15, D-24118 Kiel, Germany\\
}

\date{Accepted XXX. Received YYY; in original form ZZZ}

\pubyear{2021}

\begin{document}
\label{firstpage}
\pagerange{\pageref{firstpage}--\pageref{lastpage}}
\maketitle

\begin{abstract}
An extensive catalog of spatially-resolved galaxy rotation curves and multi-band optical light profiles for 1752 observed spiral galaxies is assembled to explore the drivers of diversity in galaxy structural parameters, rotation curve shapes, and stellar mass profiles. 
Similar data were extracted from the NIHAO galaxy simulations to identify any differences between observations and simulations. 
Several parameters, including the inner slope $\mathcal{S}$ of a rotation curve (RC), were tested for diversity.
Two distinct populations are found in observed and simulated galaxies;
(i) blue, low mass spirals with stellar mass M$^{\star}\lesssim 10^{9.3}$ M$_{\odot}$ and roughly constant $\mathcal{S}$, and (ii) redder, more massive and more diverse spirals with rapidly increasing $\mathcal{S}$. 
In all cases, the value of $\mathcal{S}$ seems equally contributed by the baryonic and non-baryonic (dark) matter. 
Diversity is shown to increase mildly with mass.
Numerical simulations reproduce well most baryon-dominated galaxy parameter distributions, such as the inner stellar mass profile slope and baryonic scaling relations, but they struggle to match the full diversity of observed galaxy rotation curves (through $\mathcal{S}$) and most dark-matter-dominated parameters. 
To reproduce observations, the error broadening of the simulation's intrinsic spread of RC metrics would have to be tripled. 
The differences in various projections of observed and simulated scaling relations may reflect limitations of current subgrid physics models to fully capture the complex nature of galaxies. 
For instance, AGNs are shown to have a significant effect on the shapes of simulated RCs. 
The inclusion of AGN feedback brings simulated and observed inner RC shapes into closer agreement. 
\end{abstract}

\begin{keywords}
galaxies: structure -- galaxies: kinematics and dynamics -- galaxies: halos
\end{keywords}

\defcitealias{Oman2015}{O15}
\defcitealias{Oman2019}{O19}
\section{Introduction}\label{sec:Intro}

In Newtonian dynamics, the shape of a rotation curve (hereafter RC) is determined by the underlying galaxy mass distribution. 
If the baryonic mass fraction is large enough, the RC shape should also be correlated to the distribution of light throughout the galaxy. 
Therefore, RC shapes provide valuable information about the combined distribution of baryonic and dark matter (hereafter DM) in galaxies, and should correlate with various structural parameters.
RC shapes have been studied extensively, and it has long been appreciated that more massive galaxies have more steeply rising RCs \citep{b81,rubin85,kent87,1991AJ....101.1231C,broeils92,flores1993,sofue01}.
The relative fraction of DM in galaxies is also proportional to their mean surface brightness or total mass, in the sense that low surface brightness (LSB) galaxies are more DM dominated throughout the galaxy than their brighter counterparts \citep{1998ApJ...508..132D, 1999ApJ...513..561C,swaters2009,lelli13,2014RvMP...86...47C, 2015ApJ...801L..20C}.
Due to their DM dominance, the RC shapes of LSB and dwarf galaxies, with a particular interest on the inner RC rise, have yielded stringent constraints and spurred debates about the small scale physics within the $\Lambda$CDM paradigm \citep{db1996,vdb01,swaters03,2005AJ....129.2119S,2006ApJS..165..461K,swaters2009,oh11,2015ApJ...801L..20C,Oman2015,Oman2019,santos18,santos20}. 
A study of the diversity in the rise of both the RC shapes and galaxy light profiles may point to correlations with baryonic processes such as feedback \citep{santos2016} and star formation efficiency \citep{Dutton2019}. 
Structural correlations of this kind are also directly linked to models of galaxy formation and evolution \citep{1998MNRAS.295..319M,2010gfe..book.....M,c07, dutton07,strucformreview}.

It has been found that the diversity in RC shapes of low mass systems is poorly reproduced by numerical simulations of galaxies \citep[e.g.,][hereafter \citetalias{Oman2015}]{Oman2015}. 
This could indicate a limitation of current numerical simulations, whose ``subgrid'' implementations (e.g., star formation mechanisms and efficiency, supernova and/or AGN feedback, recycling of metals) fail to reproduce the true nature of (spiral) galaxies \citep{dP17}. 
Alternatively, our methods of extracting RCs from observations could be flawed in a way that mature simulations could help identify \citep[e.g., inclinations, bars, extinction, distances, etc.,][hereafter \citetalias{Oman2019}]{Oman2019}. 
Either way, the comparison of galaxy RCs between observations and simulations is bound to offer new interpretations about the structure and modeling of galaxies and its components. 
Indeed, contrary to observations, numerical simulations display rather limited variations in circular velocity profiles of all galaxies at fixed mass, especially for low-mass ``dwarf'' systems \citepalias{Oman2015}. 
Observationally, dwarf galaxy RCs are reported to show greater diversity in their global shape at fixed mass \citepalias[][]{Oman2015,Oman2019}, as captured by their inner and outer RC slopes, $dV(R)/dR$, from, e.g., piecewise fits through their RC profiles, than larger galaxies. 

For clarity, a definition of ``diversity'' is warranted. 
In previous studies, diversity has been viewed as the amount of variation in rotation curve shapes between spiral galaxies with similar maximum rotation velocity \citepalias{Oman2015}. 
Here, we consider a general definition of diversity as the variation in the intrinsic spread, or scatter, of any relation between galaxy structural parameters. 
A number of tracers of galaxy diversity are investigated below.

Ultimately, an investigation of diversity in RCs alone limits the power of studies attempting to explain and characterize the notion of diversity in galaxies; light profiles may also be informative. 
While the diversity of galaxy RCs is directly linked to the baryonic and dark matter content within a galaxy, a complete investigation of diversity must also account for the diversity in galaxy light / stellar mass profiles. 
To this end, we must examine deep photometric data, as well as simulations, to investigate diversity in the shapes of projected rotation and stellar mass profiles, as well as their related structural parameters. 
The combination of rotation curves and light profiles provides a rich framework with which to explore the various projections of galaxy diversity.

Whether galaxy RC shapes are driven primarily by the baryonic and/or DM content of a galaxy, and what structural parameters drive the diversity between observations and simulations in RC shapes and light profile shapes at fixed mass, is at the core of the so-called diversity problem \citep{Oman2015}. 
We wish to address these challenges with extensive and robust comparisons between observations and state-of-the-art numerical simulations. 
A large catalog of spatially-resolved RCs and light profiles of (spiral) galaxies will enable us to explore the physical parameters that control the diversity of RC shapes and light profiles, and test whether high resolution numerical simulations can reproduce signatures of diversity in observed galaxy structural parameters.

In such comparisons, numerical simulations will be assumed to be scatter free and yield the ``intrinsic'' scatter of a scaling relation, free from deviations due to observational errors. 
Therefore, the fair comparison between the structural data sets of observed and simulated galaxies calls for using either: 
(i) observed data whose observational errors have been fully accounted for, that can thus be compared directly with error-free simulations, and/or, 
(ii) simulated data that have been broadened according to proper error propagation, and that can thus be compared directly with direct observations. 
In this paper, we explore the first method. 

The current study was largely motivated by the claim of RC diversity by \citetalias{Oman2015} and \citetalias{Oman2019} who used spatially-resolved RCs for their study of 304 spiral galaxies. 
For our re-examination of galaxy diversity, we have combined the PROBES (``Photometry and Rotation curve OBservations from Extragalactic Surveys'') catalogue of 1396 spiral galaxies with spatially-resolved extended RCs and DESI-LIS light profiles in $g$-, $r$-, and $z$-bands \citep{2019ApJ...882....6S, 2021ApJ...912...41S}
with its extension for smaller surveys called PROBES-II (Frosst et al., in prep.). 
PROBES-II includes 631 RCs of which 356 also have at least three bands of optical ($g$-, $r$-, $z$-, and/or $i$-band) photometry. 
The sum of PROBES and PROBES-II offers a nearly six-fold increase over the sample of \citetalias{Oman2015}. 
The broad range of galaxy masses covered by these samples offers rich investigations and extensive comparisons with state-of-the-art numerical simulations.
Our approach below is to first revisit galaxy diversity with RCs alone and then include multi-band light profiles to expand our discussion of diversity to stellar mass profiles. 

This paper is structured as follows. 
Our observational data are presented in \sec{ObsData}, while the NIHAO (Numerical Investigation of a Hundred Astrophysical Objects) simulations \citep{2015MNRAS.454...83W}, needed for the detailed comparisons with the observed data, are introduced in \sec{SimData}.
With these data in place, we measure RC shapes, specifically the inner RC slopes, calculated from a piecewise function and reported as $\mathcal{S}$, in \sec{MeasureingRC}. 
The uniform photometry extracted from the DESI survey for PROBES and PROBES-II galaxies is discussed in \sec{Params}. 
Our study of diversity based on observed RC slopes begins in \sec{Results}, while the ability of NIHAO simulated galaxies to match these data is discussed in \sec{simulated_RCs}. 
Insights into the drivers of RC diversity are also offered. 
Our exploration of diversity based on galaxy light and stellar mass profiles is presented in \sec{MassProfs}. 
Our conclusions are presented in \sec{Conclusions}. 
These additional dimensions into the overall paradigm of galaxy diversity are then augmented with multi-dimensional stellar mass - velocity projections in \apdx{MVDiagram}.

\section{Observational Data}\label{sec:ObsData}

As noted above (\sec{Intro}), this study used the PROBES and PROBES-II observational datasets which provide deep spatially-resolved RCs and light profiles for 1752 distinct galaxies. 
Only late-type galaxies (LTGs) with moderate inclinations ($30^\circ < i < 80^\circ$) for the measurement of RCs were included in our study. 
Moreover, both PROBES and PROBES-II have multi-band light profiles derived in the $g$-, $r$-, $z$-band from the Dark Energy Sky Instrument -- Legacy Imaging Survey\footnote{In what follows, the acronym DESI is meant to represent the Dark Energy Sky Instrument Legacy Imaging Survey (DESI-LIS).}\citep[DESI-LIS;][]{dey19}, allowing for accurate and uniform comparisons between both datasets. 

The PROBES data set is an amalgamation of seven (relatively large) surveys in the Northern and Southern hemispheres for a total of 1396 disk galaxies with available DESI photometry and extended (mostly H$\alpha$) rotation curves 
\citep{2019ApJ...882....6S, 2021ApJ...912...41S}. 
PROBES light profiles were derived uniformly via the {\scriptsize AUTOPROF} pipeline, following methods described in \cite{Stone21} and \cite{arora2021manga}; see \sec{Params} for more details.

The PROBES-II compilation assembled for this study, and presented in Frosst et~al. (in prep.), is largely based on an expansion of the RC compilation for low mass galaxies by \citetalias{Oman2015}.
We have endeavoured to expand upon that sample by including all samples of spatially-resolved RCs for low mass (dwarf + LSB) galaxies known to us. 
The PROBES-II compilation was originally meant to address a demand for predominantly lower mass galaxies, however more massive galaxies were also included. 
In addition to spatially-resolved RCs, 
PROBES-II galaxies have $g$-, $r$-, and $z$-band surface brightness profiles from DESI images, as well as $g$-, $r$-, and $i$-band surface brightness profiles from the Sloan Digital Sky Survey DR7 \citep[hereafter SDSS;][]{2009ApJS..182..543A}; again, see \sec{Params} for more details.
Only PROBES-II galaxies with both light profiles and RCs were retained for this study.
Mergers or otherwise disrupted systems, or galaxies with corrupted photometry (i.e., a star in the line of sight, bleeding, significant cosmic ray interference, failure of light profile derivation convergence etc.) were discarded. 
Altogether, the PROBES-II data set includes 631 separate LTGs from 25 different compilations, spanning all LTG Hubble types. 
356 PROBES-II galaxies have both rotation curves and at least three bands of photometry either $g$, $r$, and $z$ from DESI (N$_{\text{DESI}} = 317$) or $g$, $r$, and $i$ from SDSS DR7 (N$_{\text{SDSS}} = 39$).

\Table{RCsource} gives a broad overview of the range of parameters available for the PROBES and PROBES-II data sets.
This table presents the number of RCs recovered from each source, as well as the methods used to obtain them. 
The maximum velocity range and dynamical mass range of each sample are also presented. 
All possible PROBES, PROBES-II, and NIHAO data are included; the NIHAO data will be introduced in \sec{SimData} below. 

The PROBES and PROBES-II data sets both include the SPARC sample \citep{Lelli16}, allowing for a cross-correlation of RC and light profile measurements.
Based on the small overlap between PROBES and PROBES-II, it is inferred that structural parameters between the two observational datasets differ by at most 5\% (owing primarily to different light profile extraction techniques, as discussed in \sec{Params}).

\begin{table*}
\caption{Basic properties and parameter distributions of sources obtained for this study. 
$N$ is the number of RCs obtained from each source.
The Observation Type in the second column specifies the spectral feature(s) used to construct the RCs. 
The final three columns give the range of galaxy parameters for V$_{\text{max}}$, log(M$_{\text{dyn}}$/M$_{\odot}$), and log(M$^{\star}$/M$_{\odot}$).}
\renewcommand{\arraystretch}{0.7}
\resizebox{\textwidth}{!}{
\centering
 \begin{tabular}{| c || c | c | c | c | c |}
 \hline
 \textbf{Source} & \textbf{N} & \textbf{Observation Type} & \textbf{V$_{\text{max}}$ Range} & \textbf{log(M$_{\text{dyn}}$/M$_{\odot}$) Range} & \textbf{log(M$^{\star}$/M$_{\odot}$) Range}\\
 \hline
 \hline
 \multicolumn{6}{|c|}{\textbf{PROBES-II}} \\
 \hline
 \hline
 \citet{2014ApJ...789...63A} & 7 & H$\beta$ + [OIII] & 76.8 - 111.0 & 9.5 - 10.2 & 8.1 - 9.9\\
 \hline
 \citet{2003NewA....8..267B}& 1 & \HI & 19.0 &  8.0 & 6.55\\
 \hline
 \citet{broeils92} & 12 & \HI & 48.2 - 301.0 &  9.3 - 11.1 & 7.0 - 10.8\\
 \hline
 \citet{db2002} & 26 & H$\alpha$ & 49.7 - 112.7 &  9.0 - 10.7 & 8.0 - 9.9\\
 \hline
 \citet{2004ApJ...608..189D} & 1 & \HI & 52.0 &  9.5 & - \\
 \hline
 \citet{2008AJ....136.2648D} & 19 & \HI & 76.8 - 211.6 &  10.0 - 11.1 & 9.6 - 10.9\\
 \hline
 \citet{2008MNRAS.390..466E} & 97 & H$\alpha$ & 42.0 - 567.0 &  8.2 - 12.1 & 7.9 - 13.0\\
 \hline
 \citet{2005MNRAS.362..127G} & 24 & H$\alpha$ & 106.4 - 330.9 &  9.9 - 11.3 & 8.9 - 11.3\\
 \hline
 \citet{kauff15} & 106 & H$\alpha$ + $\sigma_{abs}$ & 87.8 - 317.9 &  9.4 - 11.4 & 7.0 - 11.7\\
 \hline
 \citet{2012MNRAS.420.2924K} & 12 & \HI & 48.7 - 85.3 &  9.3 - 9.9 & 6.9 - 8.3\\
 \hline
 \citet{2008ApJ...676..920K} & 9 & H$\alpha$ & 76.8 - 146.0 &  8.3 - 10.5 & 7.5 - 10.6\\
 \hline
 \citet{sparc16} & 175 & \HI + H$\alpha$ & 17.8 - 305.0 &  7.5 - 11.7 & 6.0 - 11.9\\
 \hline
 \citet{martinsson2013} & 30 & [OIII] & 75.3 - 214.4 &  9.8 - 11.4 & 9.1 - 11.0\\
 \hline
 \citet{2001AJ....122.2381M} & 36 & \HI + H$\alpha$ & 60.8 - 145.4 &  9.6 - 10.6 & 8.8 - 10.23\\
 \hline
 \citet{2007MNRAS.376.1513N} & 1 & \HI & 165.0 &  10.4 & 11.1\\
 \hline
 \citet{oh11} & 7 & \HI & 39.5 - 80.0 &  8.7 - 10.2 & 8.0 - 9.3\\
 \hline
 \citet{2015AJ....149..180O} & 26 & \HI & 50.0 - 211.6 &  9.5 - 11.1 & 7.0 - 10.9\\
 \hline
 \citet{1999ApJ...523..136S} & 40 & H$\alpha$ + CO & 35.9 - 266.0 &  8.8 - 11.6 & 8.0 - 11.3\\
 \hline
 \citet{2003PASJ...55...59S} & 12 & CO & 35.9 - 331.0 &  8.8 - 11.6 & 8.0 - 11.3\\
 \hline
 \citet{2000ApJ...531L.107S} & 5 & H$\alpha$ & 110.0 - 154.8 &  10.3 - 10.9 & - \\
 \hline
 \citet{swaters03} & 15 & H$\alpha$& 98.0 - 130.7 &  10.5 - 10.7 & 9.1 - 9.9\\
 \hline
 \citet{swaters2009} & 62 & \HI + CO & 35.9 - 91.5 &  8.8 - 10.4 & 8.0 - 9.2\\
 \hline
 \citet{trachternach2009} & 11 & \HI & 24.9 - 73.7 &  8.5 - 10.0 & 7.4 - 8.4\\
 \hline
 \hline
 \multicolumn{6}{|c|}{\textbf{PROBES}} \\
 \hline
 \hline
 \citet{c97} & 296 & H$\alpha$ & 45.0 - 350.0 &  8.5 - 10.0 & 7.4 - 8.4\\
 \hline
 \citet{2000ApJ...544..636C} & 171 & H$\alpha$ & 120.1 - 437.2 &  10.5 - 12.1 & 7.0 - 11.5\\
 \hline
 \citet{dale1999} & 522 & H$\alpha$ & 29.4 - 468.3 &  7.9 - 12.1 & 6.4 - 11.5\\
 \hline
 \citet{sparc16} & 175 & \HI~ + H$\alpha$ & 17.8 - 305.0 &  7.5 - 11.7 & 6.0 - 11.9\\
 \hline
 \citet{mathewson1992} & 744 & \HI + H$\alpha$ & 57.5 - 459.4 &  9.4 - 12.0 & 5.8 - 11.4\\
 \hline
 \citet{mathewson1996} & 1216 & H$\alpha$ & 24.2 - 532.8 &  8.5 - 12.0 & 6.0 - 11.5\\
 \hline
 \citet{o17} & 44 & H$\alpha$ & 9.5 - 398.5 &  6.5 - 11.0 & 6.1 - 10.7\\
 \hline
 \hline
 \multicolumn{6}{|c|}{\textbf{NIHAO}} \\
 \hline
 \hline
 \citet{2015MNRAS.454...83W} & 91 & - & 28.5 - 468.3 &  9.2 - 12.2 & 4.4 - 11.3\\
 \hline
 \hline
 \multicolumn{6}{|c|}{\textbf{NIHAO - AGN}} \\
 \hline
 \hline
 \citet{blank2019} & 18 & - & 83.4 - 249.2 &  9.0 - 10.2 & 9.0 - 11.6\\
 \hline
 \end{tabular}}
\label{tab:RCsource}
\end{table*}

\section{Simulations}\label{sec:SimData}

An inherent challenge for numerical simulations of galaxies is to replicate the observed diversity in galaxy RCs and light profiles. 
To this end, we have also investigated the properties of simulated galaxies from the NIHAO (Numerical Investigation of a Hundred Astrophysical Objects) project \citep{2015MNRAS.454...83W}.
NIHAO consists of about 100 cosmological hydrodynamical zoom-in simulations of isolated galaxies. 
The baryonic masses of NIHAO galaxies range from $10^5$ to $10^{11}$ $M_{\odot}$, providing a fair representation of galaxy types from dwarf sizes to Milky Way analogues (see \Table{RCsource} for comparison).
NIHAO simulations rely on the \textsc{Gasoline2} code in a flat $\Lambda$CDM cosmology \citep{2017MNRAS.471.2357W}. 
The resolution of the simulations scales with halo mass such that all simulations have a similar numbers of particles of the order of $10^6$; this implies that even the lowest mass galaxies are well resolved. 

Stars were allowed to form  when gas particles reach $T<15000$ K and a density of $n \geq 10.3$ particles per cubic centimeter.
Supernovae were simulated following the blast-wave formalism \citep{2006MNRAS.373.1074S}, with the addition of stellar feedback in the form of winds from massive stars before they explode, the so-called ``Early Stellar Feedback'' \citep{2013MNRAS.436..625S}. 

A subset of the NIHAO galaxies were re-simulated with the addition of AGN feedback \citep{blank2019}. 
Supermassive black holes (SMBH) were seeded in all (central) haloes with a virial mass above $10^{10}$ M$_{\odot}$, the 
gas accretion onto the BH is regulated by the Bondi-Hoyle-Lyttleton accretion, and feedback is purely thermal \citep[see][for more details]{blank2019}.

In order to single out the possible effects of an AGN on the rotation curves, we will present results separately for the original NIHAO set \citep{2015MNRAS.454...83W} as well as the additional 18 galaxies (with virial masses, M$_{200} > 10^{12}$ M$_{\odot}$) evolved with AGN feedback (hereafter ``NIHAO-AGN'').
NIHAO galaxies with stellar masses below $10^{7}$ M$_{\odot}$ were discarded as they lie below the mass range of our observed galaxies, leaving us with a total of 69 NIHAO galaxies plus the NIHAO-AGN sample, for a total of 87 simulated galaxies. 

NIHAO galaxies have been shown to match a number of observed galaxy properties and scaling relations \citep{2016MNRAS.463L..69M, 2016ApJ...824L..26O, 2017MNRAS.468.3628B, d17}. Previous related studies by \cite{santos18} and \cite{santos20} have also suggested that the NIHAO simulated galaxies generally match the observed RC diversity reported by \citetalias{Oman2015}. 
Below, we use the same NIHAO simulations to verify and expand upon these studies, using an extended observed data set as well as alternative methods to characterize the {\it diversity} of the observed and simulated RCs and light profiles. 

For both NIHAO and NIHAO-AGN galaxies, the RCs were calculated using the circular velocity (using the midplane gravity from the assumed flat disk) as function of galactocentric radius making use of the \pynbody~python package \citep{Pontzen2013}. 
Velocities were measured perpendicular to the angular momentum vector of the stellar particles, thus removing a possible source of scatter due to inclination corrections.
The simulated RCs, as well as the stellar mass, gas mass, and stellar mass surface density profiles, were all traced out to a radius of $0.2$R$_{200}$. 
We found little difference between RCs derived via the true galaxy potential (V$^{\text{pot}}$, in the disc plane) or assuming a spherical potential (V$^{\text{sph}}$, V$^2$ = GM/R). 
Indeed, galaxy RCs derived by tracing the galactic circular potential (V$^{\text{pot}}$) and those derived by tracing the circular velocity ($V^{\text{sph}}$) have
 $V^{\text{pot}}_{\text{2kpc}} / V^{\text{sph}}_{\text{2kpc}} \geq 0.76$, with a mean of 0.91, and $V^{\text{pot}}_\text{last} / V^{\text{sph}}_\text{last} \simeq 0.99$, matching the results of \citet{santos18}, where V$_{\text{2kpc}}$ denotes the rotational velocity measured at a radius of 2 kpc, and V$_{\text{last}}$ indicates the last measured rotational velocity point on the RC. 

In what follows, we stress that statistical completeness is by no means achieved for both our observed and simulated data sets which all suffer their own limitations. 
While our broad coverage of observed and simulated data offers some averaging of sample differences and internal biases, all comparisons below must still be treated with care. 

\section{Measuring Rotation Curves}\label{sec:MeasureingRC}

Numerous methods to characterize RC shapes exist \citep{Brandt60,rubin85,madore1987,Bertola1991,1991AJ....101.1231C,flores1993,Giovanelli1994,c97,lelli13,kauff15,Oman2015,Sofue2016}, ranging from fitting polynomial functions \citep{Bertola1991,Giovanelli1994, c97,lelli13,kauff15}, to computing the scatter of points on the V$_{\text{2kpc}}$ - V$_{\text{last}}$ relation \citepalias[e.g.,][]{Oman2015}. 
Each method, if used and interpreted appropriately, can adequately represent some aspects of the shape of the rotational velocity of a galaxy. 
For comparison with \citetalias{Oman2015}, we have focused on the inner RC slope and investigated a variety of measurement techniques\footnote{Some of these techniques involved linear fits to the RC up to fixed or isophotal radii, linear fits to some fraction of radial extent of the RC, linear fits to the RC to some fraction of the total velocity, polynomial fits, etc.}.
The piecewise linear function fit to the RC using maximum likelihood estimation was ultimately adopted to represent the inner and outer RC slopes of each observed and simulated galaxy RC in our compilation.
For simplicity, the inner RC slope will be referred to as $\mathcal{S}$ (see \Eq{piecewiseRC}). 

\subsection{Fitting Rotation Curves}\label{sec:rc_fit}

RCs follow simple shapes; starting at the center of a galaxy with a rest-frame velocity of zero, the bulge rotates as a solid-body with $V(r) \propto r$, before transitioning into a disk with $V(r) \propto V_c$, where $V_c$ approximates some constant rotational velocity.
RCs may also show declining profiles in their outskirts (found for 14\% of the PROBES and PROBES-II galaxies RCs). 
An RC can also be fitted with numerical functions that match most of its features with only a few fitting parameters \citep{Bertola1991,c97,2002ApJ...571L.107G}.
Such models are useful for measuring the shape of the curve, and extrapolating RCs beyond the last observed velocity data points if necessary. 
Needless to say, extrapolations require genuine care. 

For this study, we have adopted in order of increasing complexity two RC fitting functions: the piecewise (PW) linear model and the multi-parameter (MP) model 2 of \cite{c97}. 
The PW model is largely used to measure the inner and outer slopes of each RC, while the MP model is more sensitive to RC nuances such as sudden velocity fluctuations from a bulge. 
Both the PW and MP models are fit to observational and simulated RCs via maximum likelihood parameter estimation.
Errors on the fit parameters were estimated with a Markov Chain Monte Carlo (MCMC) method.

In some cases, individual galaxies have multiple RCs obtained from separate authors in the combined PROBES and PROBES-II compendium. 
When this occurs, we stack the common RCs, regardless of velocity tracer or the method of derivation.
When a RC has been obtained unfolded, we first fit the \citet{c97} model (see \sec{MPmodel}) and then subtract the systematic velocity, V$_{0}$, and the x-axis offset, $r_0$, such that the RC passes through (0,0). 
We then fold the RC so that all radii and velocities values are along the positive axes of R and V. 
The final stacked RCs are hereafter used exclusively. 
No significant differences in RC shape are found for different velocity tracers.

\subsubsection{The Piecewise Function and Calculating $\mathcal{S}$}\label{sec:rcslope}

We present a well-defined, easy to implement, method to measure the RC's inner slope. 
As the inner RC can be characterized by solid body rotation, a simple model choice is a straight line.
While the outer rotation curve may have more structure, it is not the primary focus of this study and we also model it as a straight line.
Such a model is idealised by a piecewise linear fit, where the inner and outer slopes intercept at the same location within the RC. 
The slope, $\mathcal{S}$, of the fit to the inner part of the RC with our piecewise function is used to explore RC diversity below.

\begin{figure*}
	\includegraphics[width=\textwidth]{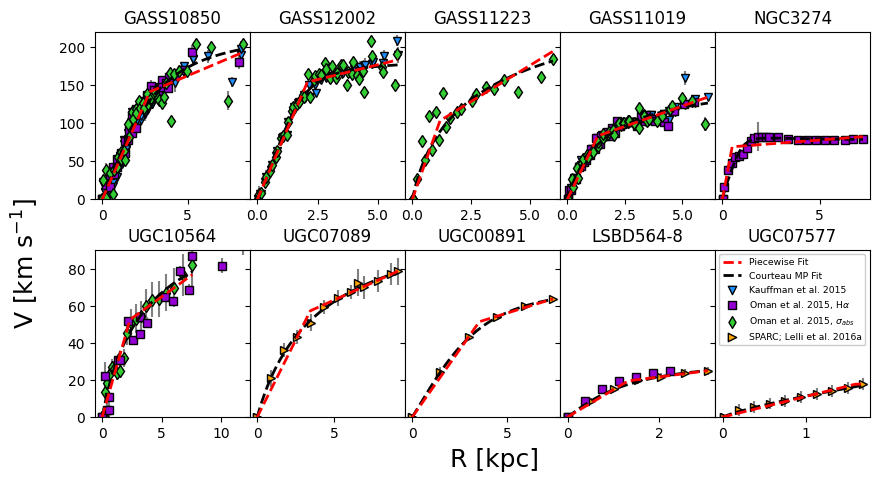}
    \caption{PROBES-II RCs of varying \protect V$_{\text{max}}$ with MP and piecewise (PW) fits for ten representative galaxies, named above each subplot.
    The RC data are represented by coloured points, the PW fits are shown as red dashed lines, and the \protect\cite{c97} MP model is shown as a blacked dashed line. 
    Blue inverted triangles show data from \protect\cite{kauff15}, purple squares are \protect H$\alpha$ measurements compiled by \protect\citetalias{Oman2015}, orange triangles are from \protect\cite{sparc16}, and light green diamonds show RC data from stellar absorption lines from \citetalias{Oman2015}.
    }
	\label{fig:TLmodel_example}
\end{figure*}

A piecewise fit allows for multiple (here two) linear fits to the data over some radial range. A transition radius, R$_{pw}$, appears where the linear fits meet. 
We assume that the piecewise RC fit is continuous.
Therefore, 

\begin{equation*}
V =  \begin{cases} 
      \mathcal{S} R + b_1 & x\leq R_{pw} \\
      \mathcal{S}_{outer} R + b_2 & x > R_{pw} 
  \end{cases},
\end{equation*}
where $b_1 = 0$ for the fit to pass through the origin. 
At the transition radius,
\begin{eqnarray}
    \mathcal{S} R_{pw} &=& \mathcal{S}_{outer} R_{pw} + b_2\\
    b_2 &=& R_{pw} (\mathcal{S} - \mathcal{S}_{outer}).
\end{eqnarray}
Substituting into the first part of the piecewise function yields:
\begin{equation}\label{eq:piecewiseRC}
V =  \begin{cases} 
      \mathcal{S} R & R\leq R_{pw} \\
      \mathcal{S}_{outer} R + R_{pw}(\mathcal{S} - \mathcal{S}_{outer}) & R > R_{pw} 
  \end{cases},
\end{equation}  
  
resulting in a piecewise linear fit that is continuous at $R_{pw}$. 
This model provides a simplified representation of the RC inner and outer slopes. 
Examples of PW fits for a random sub-sample of PROBES-II galaxies are shown in \Fig{TLmodel_example}.

\subsubsection{The Courteau 1997 Multi-Parameter Model}\label{sec:MPmodel}

The \cite{c97} MP formulation for fitting spatially-resolved galaxy rotation curves is given by: 
\begin{equation}\label{eq:MP}
    V(r) = V_0 + V_c \frac{(1+x)^\beta}{(1+x^\gamma)^{1/\gamma}},
\end{equation}
where $x = 1/R = r_t/(r - r_0)$. 
The parameters $r_0$ and $v_0$ are the location and heliocentric velocity of the galaxy center, respectively; 
$r_t$ is the transition radius from a rising to a flat curve,
and $v_c$ is an asymptotic velocity. 
The parameter $\beta$ allows for a declining or rising RC in the galaxy's outskirts. 
The parameter $\gamma$ changes the sharpness of the turnover from the inner to the outer parts of the galaxy, and can be especially useful for modeling dominant bulges.  
Given its multiple degrees of freedom, the multi-parameter (MP) model frequently provides accurate RC fits over a broad mass range. 
For this reason, the MP function (Eq.~\ref{eq:MP}) is our model of choice for RC interpolations or extrapolations; the measurement of inner and outer RC slopes will still rely on the PW model.

\subsubsection{RC Model Performance}

All PROBES, PROBES-II, NIHAO, and NIHAO-AGN RCs were fitted with the MP and PW models. 
Visual inspection was performed on all fitted RCs to ensure accurate modelling. 
Additionally, $\chi^2$ values for each fit were computed, with the PW and MP models performing equally well.
For quality control, the velocity V$_{23.5}$ at R$_{23.5}$ (the radius at which the surface brightness reaches a value of $23.5$ $i$-mag arcsec$^{-2}$) was linearly interpolated for the PROBES-II RC data set and compared to the estimated values from the various models; essentially performing a K-fold cross validation of the models \citep{Hastie2009}. 
The MP and PW models identified log(V$_{23.5}$) equally well with a small rms scatter of $\sigma = 0.1$ between the interpolated log(V$_{23.5,\text{MP}}$) and log(V$_{23.5,\text{PW}}$). 
This test lends support for PW results given the reliability of the MP model. 
Similar results were found for other velocities, such as V$_{\text{2kpc}}$. 

We pursue our exploration of the RC diversity with the PW model. 
An example of a variety of PROBES-II RC fits with the PW model is found in  \Fig{TLmodel_example}. 
While the PW model simplifies the observed RC, it follows the general variations of the inner and outer parts of each RC. 

The PW model can occasionally struggle to correctly identify the inner and outer slopes of large galaxies, particularly when for hosts of a massive bulge. 
In such cases, the PW model is rejected (occurring less than 4\% for our observed RCs, and about 10\% for the simulated RCs), and a modified MP model fit is used instead. 
The inner slope fitting procedure is modified for those systems by fitting a line between the RC origin and the last point before the transition radius identified by the MP Model, while the outer slope is found by fitting all points beyond the transition radius. 
The most notable difference between this technique and the PW model is that the inner and outer slopes occasionally do not meet within the RC. 
In such cases, the associated inner and outer slopes do not wholly represent the shape of the RC. 
However, when considered individually, the inner and outer slopes still represent the slopes of the various portions of the RC. 

\subsection{Errors of Rotation Curve Slopes}

The RCs collected in this study have been corrected for various effects (e.g., inclination, flux calibration, beam smearing effects, asymmetric drift, etc.).
Still, various effects contribute to errors in measured RC slopes with the PW model. 
Following \citet{lelli13}, these include errors from the fit, errors from inclination corrections, and errors from the distance measurements.

These errors are added in quadrature: 
\begin{equation}\label{eq:slope_err}
    \delta \mathcal{S} = \sqrt{(\delta_s)^2 + \left( \mathcal{S} \frac{\delta_i}{tan(i)}\right)^2 + \left( \mathcal{S} \frac{\delta_D}{D} \right)^2},
\end{equation}
where $\mathcal{S}$ is the inner slope, $\delta_s$ is the error on the fitted slope, as estimated from the posterior probability distribution from an MCMC, $\delta_i$ is the error on the fitted inclination $i$, and $\delta_D$ is the error on the distance $D$. 
The distance error term is often negligible if distances come from TRGB or Cepheid techniques (which have $\Delta$D/D$\sim$5\%); however, they can be significant for Tully-Fisher estimates (with $\Delta$D/D$\sim$25\%). 
By influencing the inferred size of galaxies, distance errors may yield incorrect RC slopes.
Ultimately, for nearly all observed galaxies, the error on the RC inner slopes is negligible when calculated with \Eq{slope_err}, averaging roughly $\langle \delta \mathcal{S} \rangle \sim 5$ km s$^{-1}$ kpc$^{-1}$, or $\langle \log(\delta \mathcal{S} / $km s$^{-1}$ kpc$^{-1}) \rangle \sim 0.04$. 
The error analysis for $\mathcal{S}$ above is almost certainly a lower bound: the mixing of RCs from various sources and methods (e.g., H$\alpha$, \HI, and CO profiles) all affect the measurements of RC slopes.

\subsection{Structural Parameters from Rotation Curves}
Finally, 
a variety of structural parameters are derived from our RCs. 
Before RCs are fit with any model, standard inclination corrections \citep[e.g.][]{c97} are applied, if not already done so by the original authors. 
Our deprojection of the RCs uses the inclination of the last isophote from the \textit{r}-band photometry, defaulting to DESI isophotal inclinations when available. 

For the extraction of specific velocity values, we fit RCs with the MP model and derive velocity values from the fitted function. 
For instance, the maximum rotational velocity, V$_{\text{max}}$, and the rotational velocity at the radius of the 23.5~\magss isophote, V$_{23.5}$, are derived from the fitted MP model (\sec{MPmodel}). 

\section{Measuring Light Curves}\label{sec:Params}

With our RC data in place, we now describe the methods used to extract photometric parameters from our observational data. 
The extraction of structural parameters for the PROBES and PROBES-II samples follows the methods presented in \cite{arora2021manga}, \cite{Stone21}, and Frosst et al. (in prep.).

The (interactive) {\scriptsize{XVISTA}}\footnote{Developed at Lick Observatory, the {\scriptsize XVISTA} image reduction and analysis package is maintained by Jon Holtzman at NMSU (New Mexico State University); http://ganymede.nmsu.edu/holtz/xvista/.
The {\scriptsize XVISTA} photometry package offers a wide range of image processing tools used in this study.} 
ellipse fitting program {\scriptsize{PROFILE}}  \citep{1985MNRAS.216..429L,c96} was used to extract SB profiles for PROBES-II galaxies, while the photometry for PROBES galaxies relied on the automated isophotal fitting software {\scriptsize{AUTOPROF}} \citep{Stone21}. 
The use of any one of these isophotal fitting softwares is inconsequential to the resulting light profiles and the extracted structural parameters, though {\scriptsize{AUTOPROF}} can extend SB profiles to light levels roughly 2 magnitudes deeper \citep{arora2021manga,Stone21}. 
More details about galaxy surface photometry based on {\scriptsize XVISTA} can be found in  \cite{1985MNRAS.216..429L,c96,McDonald2011,hall12,Gilhuly2018}. 

The photometric reductions involved a number of steps outlined below.
The DESI photometry\footnote{https://www.legacysurvey.org} (in the $g$, $r$, and $z$ bands) was available for most sources, and the $r$-band images were preferred (for the choice of isophotal contours) for their stable sky background, high signal-to-noise ratio, and low sensitivity to dust extinction. 
The $r$-band is also the reddest common band between SDSS and DESI photometry ($z$-band SDSS imaging exists but its lower signal-to-noise ratio is unfavorable for this study). 
Indeed, if DESI photometry is not available for a galaxy, SDSS photometry was used instead (roughly ${\sim}9$\% of the PROBES-II sample, and no PROBES galaxies). 
The isophotal solution determined for the $r$-band image could then be applied directly onto the other $g$-band and $i$-band images (for SDSS photometry) or $g$-band and $z$-band images (for DESI photometry). 
This ensured uniform colour gradients, since the measured pixels used to calculate the SB profiles in each image remain the same.
 
From our multi-band profiles, various structural parameters may be derived, such as luminosities, colours, surface mass densities, and others \citep{arora2021manga}. 
Stuctural parameters, such as magnitudes and surface brightnesses, must be corrected for Galactic extinction ($A_g$), geometric projections ($\gamma$), internal extinction ($A_i$), and k-correction ($A_k$)  \citep{c96,hall12}.  
For instance, apparent magnitudes for each band were corrected as:
\begin{eqnarray}
    m_{corr,\lambda} = m_{obs,\lambda} - A_{g,\lambda} - A_{i,\lambda},
\end{eqnarray}
where $\lambda$ denotes the band of interest. 
Galactic extinction corrections for all PROBES and PROBES-II galaxies were taken from the NASA/IPAC Extragalactic Database (NED), which themselves follow \cite{Schlafly2011}. 
Cosmological k-corrections are minimal and ignored given the proximity of PROBES and PROBES-II galaxies.

The internal extinction and geometric projection corrections require greater care, in part because of their morphological (Hubble T Type) dependence.
Our corrections involved a forward least-squares linear fit between the log of the cosine of the inclination of the galaxy, and the structural parameters of interest. 
The correction for any structural parameter, $X$, is therefore modeled as
\begin{eqnarray}\label{eq:extinction_corr}
    \log X_0 = \log X_i + \gamma\log(\cos(i)),
\end{eqnarray}
where $X_0$ is the corrected parameter, $X_i$ is the observed projected parameter, $i$ is the inclination of the galaxy, and $\gamma$ is the correction factor. 
This follows the correction scheme of \cite{Giovanelli1994}, further explored in \cite{2021ApJ...912...41S} and \cite{arora2021manga}. 
Across samples and parameters, correction factors can vary widely, particularly with small data sets.
To circumvent this predicament and achieve higher accuracy, our correction used the combined PROBES-II and PROBES data sets, as well as a subset of the extensive compilation of light profiles from DESI for the MaNGA survey by \cite{arora2021manga}. 
The combined data set of 3544 galaxies was divided into three Hubble T Type categories: (i) Types 0 - 3, (ii) Types 4 - 5, and (iii) Types 6 - 10. 
Only galaxy magnitudes and sizes were corrected. 

Two methods could be used to infer inclination-corrected colours. 
Firstly, raw colours may be calculated and the resulting term is corrected by testing against inclination according to \Eq{extinction_corr}. 
Secondly, the individual apparent magnitudes for a given colour term can each be corrected for inclination effects, and then carried through to calculate the colour. 
In both cases, one should arrive at the same distribution of corrected colours, and in principle, whether the first or second method is applied should not matter. 
For simplicity, the second method was applied as the number of calculations is reduced.
We shall return to concerns about galaxy diversity based on photometric data in \sec{MassProfs}.
However, all parameters used in the analysis below were corrected for inclination following \Eq{extinction_corr}.

\subsection{Structural Parameters from Light Profiles}
The structural parameters derived from light profiles for this study include isophotal radii, most commonly R$_{23.5}$, the radius at which the surface brightness in the \textit{r}-band reaches a value of $\mu_r = 23.5$~\magss; light concentration, C$_{28}$; and the central surface brightness, $\mu_0$.
Multi-band light profiles are used to calculate galaxy colours, e.g., (g - z); stellar mass, M$^{\star}$; and the surface mass density, $\Sigma^\star$, with the use of mass-to-light colour relations \citep[hereafter MLCR,][]{bell01,roediger15}. 

These parameters are often measured at various radii and are linearly interpolated from the light profiles. 
Where possible, the fiducial radius R$_{23.5}$ is used. 
For instance, the stellar mass, M$^{\star}_{23.5}$, is measured at R$_{23.5}$.
We also report $M^{\star}_{\text{last}}$ measured at the last recorded point on the multi-band light profile.
The surface mass density at 1 kpc is referred to as $\Sigma^\star_1$, and is calculated as M$^{\star}_{\text{1kpc}}/\pi$. 
For the central surface brightnesses, $\mu_0$, a linear interpolation of the light profile to the center ($R=0$) is used if necessary. 
Detailed explanations for the measurement of all structural parameters in this study are presented in \cite{Stone21} and Frosst et~al (in prep.) 

The total enclosed mass of the galaxy measured from the RC at the corresponding radius, is given by  
\begin{equation}\label{eq:Mtot}
  \text{M}_{\text{tot}}(R) = \frac{V_{\text{circ}}^2R}{G} = 2.33\times10^{5}R V^{2}/\sin(i)^2
\end{equation}
with $R$ in kpc and $V$ in \kms. 
Dark matter fractions are calculated as f$_{\text{DM}}(R) =$ M$_{\text{DM}}(R)/$M$_{\text{tot}}(R)$, where M$_{\text{DM}}(R)$ = M$_{\text{tot}}(R)$ - M$_{\text{bar}}(R)$. 

Gas masses were obtained from the literature for both PROBES and PROBES-II, where possible. 
If only \HI~fluxes are available, \HI~masses were computed through typical transformations \citep{haynes84}, and subsequently converted to gas masses via M$_{\text{gas}} = 1.35M_{\text{HI}}$ \citep{2016ApJ...824L..26O}.  
Gas fractions may then be calculated as 
f$_{\text{gas}} = $M$_{\text{gas}}/$M$_{\text{tot}}$. 
Baryonic masses were then evaluated as 
M$_{\text{bar}} =\ $M$^{\star} +\ $M$_{\text{gas}}$. 
Ultimately, only a small subset (9.4\%) of the PROBES and PROBES-II galaxies had available gas masses, and of those only a few had reliable uncertainties. 
Therefore, scaling relations including M$_{\text{gas}}$, f$_{\text{gas}}$, or M$_{\text{bar}}$ must be handled with care. 
Similarly, only galaxies with f$_{\text{DM}}$ for log(M$^{\star}$/M$_\odot$) > 9.5 are presented, as their gas masses are only a small fraction of the total mass
and can thus be cautiously ignored in the computation of dark matter masses. 

We report light and stellar mass concentrations as C$_{28}$ and C$^\star_{28}$ respectively. 
This is accomplished by measuring the radii at which 80\% and 20\% of the light/mass is enclosed, hereafter referred to as R$_{80}$ and R$_{20}$ respectively. 
The concentration, either by light or stellar mass, is then calculated as 
\begin{equation}
    C_{28} = 5\log\left(\frac{R_{80}}{R_{20}}\right).
\end{equation}

Finally, we compiled a list of AGN hosts within the PROBES and PROBES-II samples via a literature search in an effort to understand the influence of AGNs on RCs. 
Only 41 AGN hosts were identified in our observational samples, of which only 23 had the appropriate photometry and were within our selection criteria. 
The number of identified AGNs is undoubtedly a lower limit. 

\section{Inner RC Slopes of Observed Galaxies}\label{sec:Results} 

As discussed previously, our analysis of diversity in observed RCs relies on PW fits. 
In this section, the RC data (PROBES and PROBES-II RCs) are first investigated for their diversity across the entire mass range. 
Parameter correlations with the inner RC slope $\mathcal{S}$ will be used to quantify the drivers of RC shape variations and calculate scatters at fixed structural parameter values. 
Our presentation focuses on observed galaxies first, as many galaxies in this datatset have already been considered for RC diversity studies \citep{swaters2009,lelli13,Oman2015, kauff15, 2016MNRAS.462.3628R, santos18, santos20}, allowing a more direct comparison between methods and results. 
Scaling relations derived for observed galaxies will be revisited in \sec{simulated_RCs} using an expanded dataset that includes simulated NIHAO and NIHAO-AGN galaxies. 

In what follows, we determine the observed parameters that correlate best with $\mathcal{S}$ before assessing if simulated galaxies can replicate the slopes, intercepts, scatters, and diversity of the strongest observed relations with $\mathcal{S}$.

\subsection{Exploring High Inner Slopes}\label{sec:agnslope}
A broad range of $\mathcal{S}$ values, from $5.9$kms$^{-1}$kpc$^{-1}$ to nearly $4000$kms$^{-1}$kpc$^{-1}$, are found for the observational RCs.
The steepest $\mathcal{S}$, found at high mass, tend to deviate from the trends defined by intermediate and low mass galaxies. 
Our investigation of RC slopes therefore requires a preliminary discussion of the high inner slopes, $\mathcal{S}$, found for PROBES and PROBES-II galaxies with the PW function.
The suggestion that exceptionally high RC inner slopes could result from the methods by which RCs are measured can be discarded given that stellar and gas RCs typically yield the same velocity fields \citep{martinsson2013, Oman2015}. 
More specifically, galaxies with both \HI~and \ha~data display similar $\mathcal{S}$ values and RC shapes. 
No correlation between $\mathcal{S}$ and RC measurement / derivation methods can be identified.

\begin{figure}
	\includegraphics[width=\columnwidth]{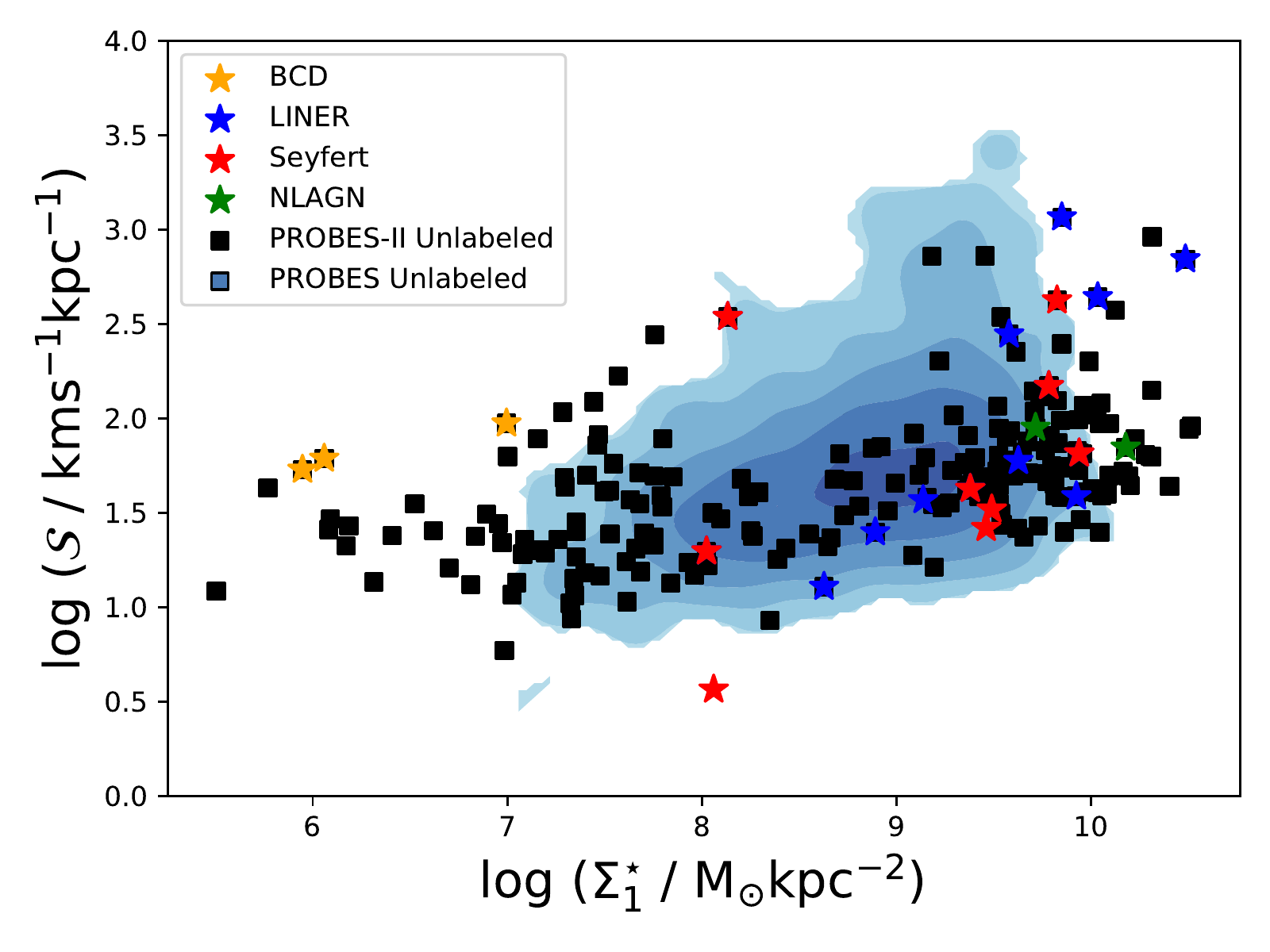}
    \caption{Inner RC log slope, plotted against $\Sigma^\star_1$, the stellar mass density within 1 kpc. 
    LINER, Seyfert, NLAGN, or BCD galaxies are labelled with coloured stars. 
    Unlabeled PROBES galaxies have blue contours, while unlabeled PROBES-II galaxies are shown as black squares. 
    Overall, the observed inner RC slopes seem unaffected by the presence of an AGN.} 
	\label{fig:AGN}
\end{figure}

One wonders if the scatter in observed RCs could result from significant disturbances in the central gas kinematics induced by the AGN \citep{2016ApJ...819..148K,Manzano_King_2020}. 
For instance, most PROBES-II galaxies with high $\mathcal{S}$ are found to host an AGN.
However, considering the combined PROBES and PROBES-II samples, there exists AGN-hosting galaxies with lower $\mathcal{S}$ values that match the observed range of non AGN-hosting galaxies with similar mass, suggesting that AGNs alone do not affect $\mathcal{S}$ appreciably. 
If AGNs were to cause diversity in the observed $\mathcal{S}$, their range of influence would therefore be broad, affecting some galaxies at fixed mass more than others.
This is exemplified in \Fig{AGN} where AGN hosting systems are labelled with coloured stars.
No observed trends in RC shape, as measured by $\mathcal{S}$, are found for AGN hosts. 
This agrees with previous observational studies reporting no significant differences between the RC shapes of AGN hosts and non-AGN hosting galaxies \citep[][and references therein]{sofue01}. 

However, the RCs of AGN-hosting galaxies may still need further investigation as current samples remain small and the lack of observed trends with $\mathcal{S}$ is at odds with numerical predictions.  
We pursue this discussion in \sec{InnerSlopesAll} where RC shapes and log($\mathcal{S}$) values are presented for simulated galaxies with and without SMBH feedback. 

\begin{figure*}
	\includegraphics[width=\textwidth]{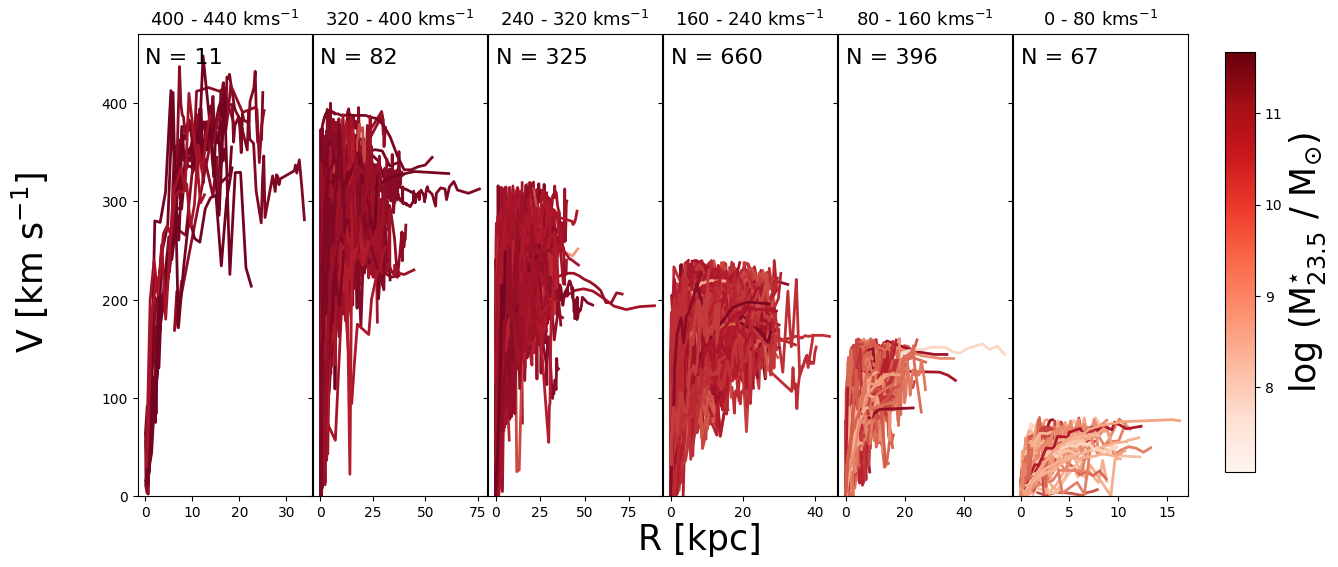}
    \caption{PROBES and PROBES-II RCs plotted in 80~\kms~bins and coloured according to stellar mass, measured at R$_{23.5}$, with whiter/redder galaxies having lower/higher stellar masses. 
    The curves are layered such that galaxies with lowest mass appear at the front of each panel. 
    The range of RC shapes is consistent across panels with similar numbers of galaxies.}
    \label{fig:RCbinneddiversity}
\end{figure*}

\subsection{Revisiting the Notion of RC Diversity} \label{sec:reaffirm_diversity}

A basic investigation of the range of RC shapes can begin with a visual comparison between RCs of similar maximum velocity, as shown in \Fig{RCbinneddiversity} for all observed RCs in V$_{\text{max}}$ bins of 80 \kms. 
Each RC is colour-coded with respect to M$^{\star}_{23.5}$. 
The scatter of RC shapes within a given velocity bin exemplifies previous notions of diversity in RC shapes at fixed stellar mass / maximum velocity. 
Visual inspection of \Fig{RCbinneddiversity}, as well as a more quantitative analysis of RCs with the PW function, both confirm the comparable scatter in RC shapes at all radii from the lowest to the highest mass bins, at odds with other claims of excess RC shape diversity at low masses \citep[e.g.,][]{Oman2015}. 

\begin{figure*}
	\includegraphics[width=\textwidth]{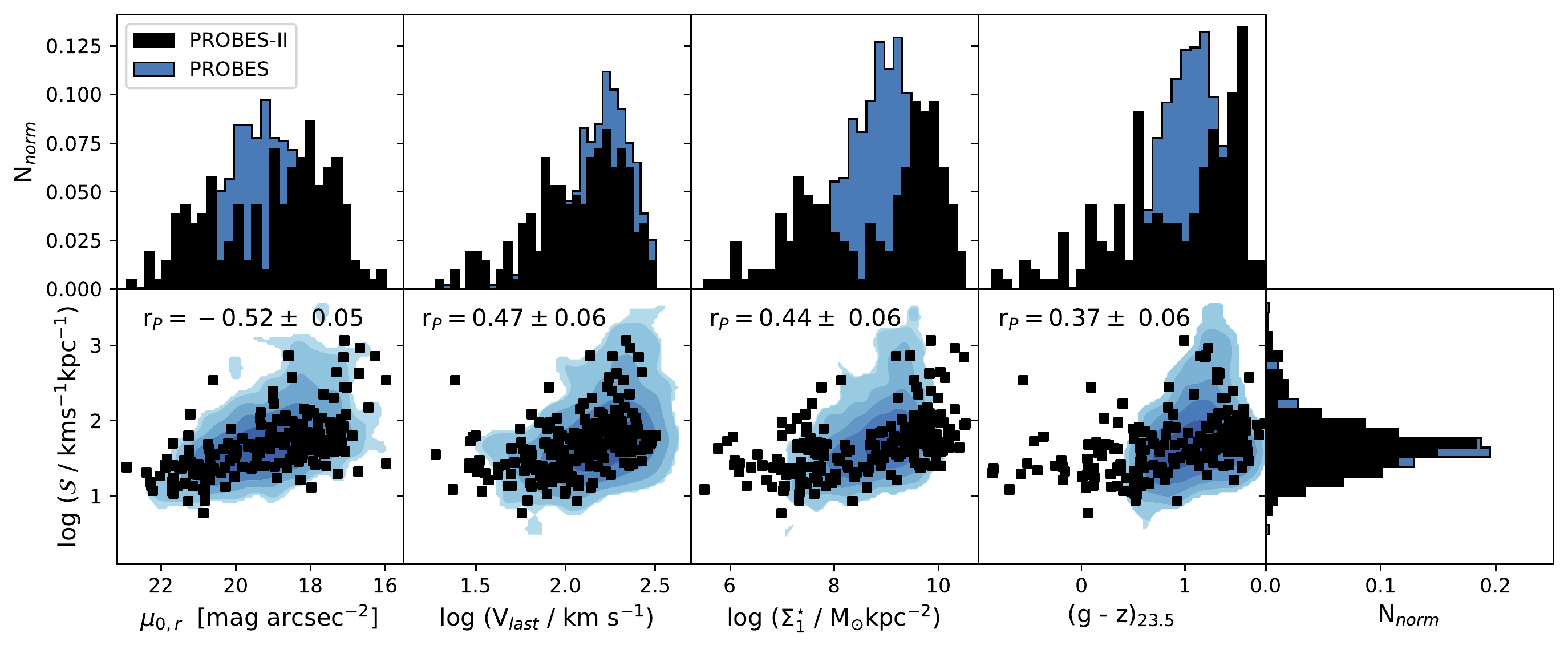}
    \caption{Structural parameters correlating most strongly (in decreasing order) with the inner log slope, $\mathcal{S}$. From left to right, $\mu_{0,r}$, log($\Sigma^\star_1$), 
    log(V$_{\text{last}}$), and (g-z)$_{23.5}$. 
    The velocities are interpolated with the MP model if necessary. 
    PROBES and PROBES-II galaxies are displayed as blue background contours and black squares, respectively. 
    The distributions of each parameter space is shown for each galaxy catalogue above each panel and the bottom right panel.
    The PROBES and PROBES-II samples have matching log~$\mathcal{S}$ distributions, and complement each other in other structural parameter distributions.}
    \label{fig:slopeSBC28}
\end{figure*}

\subsection{Inner RC Slope Correlations}\label{sec:innerslope}

Since all RCs in the PROBES and PROBES-II galaxies were fitted with the PW model (\sec{MeasureingRC}), we can examine inner RC slope measurements, $\mathcal{S}$, for all 1752 galaxies with matching photometry. 
Resulting correlations between these slopes and various galaxian structural parameters can then reveal the signatures and drivers of galaxy diversity. 
While trends with inner RC slopes have been explored before \citep{kent87,lelli13,kauff15,santos18,santos20}, our extensive data base and analysis may offer a new perspective on RC diversity. 
\Fig{slopeSBC28} was constructed with the goal of determining if inner RC diversity is primarily driven by the baryonic or DM components of a galaxy, or both. 
Many other structural parameters were tested for correlations with $\mathcal{S}$ (see \Table{Paramcorrelations}), but only those with the strongest correlations, as gauged by the Pearson linear correlation coefficient, r$_P$, are shown here. 
These are, in decreasing order, 
$\mu_{0,r}$, V$_{\text{last}}$, $\Sigma^\star_1$ and V$_{23.5}$. 
We show the (g-r)$_{23.5}$ colour, as opposed to V$_{23.5}$, to give a more holistic view of the parameter space. 
As stated above, the range of $\mathcal{S}$ in the observed samples extends from $5.9$ km s$^{-1}$ kpc$^{-1}$ to $10^{3.6}$ km s$^{-1}$ kpc$^{-1}$. 
The Pearson linear correlation coefficient, r$_P$, is shown in the top left corner of all panels in \Fig{slopeSBC28}. 

\begin{table*}
	\centering
	\caption{Overview of observed structural parameters versus $\mathcal{S}$ tested for diversity signatures for all 1752 PROBES and PROBES-II galaxies. 
	Listed are the Pearson's correlation coefficient, r$_P$, the slope, $m$, the intercept, $b$, the Bayesian intrinsic scatter, $\sigma_i$, the BP slope, and the BP test p-value.
	The quantity $f_{\text{DM}}(<R_{pw})$ is computed only for galaxies with log(M$^{\star}_{23.5}$ / M$_{\odot}$) $> 9.5$; see text for explanations.
	} 
	\label{tab:Paramcorrelations}
	\begin{tabular}{lrrrrrrc}
		\hline
		Observed Parameter & {N \ } & {r$_P$ \ } & {m \ } & {b \ } & $\sigma_i$ & BP Slope & BP p-value\\
		\hline
		\hline
		log(M$_{\text{gas}}$ / M$_{\odot}$) & 115 & 0.17 & 0.17 & 0.07 & 0.38 & 0.06 & 0.22\\
		log(M$^{\star}_{23.5}$ / M$_{\odot}$) & 1368 & 0.40 & 0.26 & -0.80 & 0.39 & 0.02 & 0.37\\
		log(M$_{\text{bar}}$ / M$_{\odot}$) & 115 & 0.42 & 0.19 & -0.21 & 0.49 & 0.02 & 0.49\\
		log(M$_{\text{DM}}$ / M$_{\odot}$) & 1368 & 0.15 & 0.25 & -1.07 & 0.40 & 0.04 & 0.09\\
		log(M$_{\text{tot}}$ / M$_{\odot}$) & 1368 & 0.16 & 0.26 & -1.11 & 0.38 & 0.05 & 0.04\\
		f$_{\text{gas}}$ & 115 & -0.26 & -1.44 & 1.75 & 0.17 & -0.37 & 0.27\\
		$f_{\text{DM}}(<R_{pw})$ & 1008 & -0.23 & 0.47 & 1.89 & 0.38 & 0.11 & 0.40\\
		(g - z)$_{23.5}$ & 1368 & 0.37 & 0.48 & 1.23 & 0.41 & 0.04 & 0.82\\
		(g - z)$_{\text{1 kpc}}$ & 1368 & 0.35 & 0.32 & 1.30 & 0.40 & 0.05 & 0.75\\
		C$_{28}$ & 1368 & 0.34 & 0.25 & 0.86 & 0.39 & 0.06 & 0.14\\
		C$^\star_{28}$ & 1368 & 0.23 & 0.14 & 1.19 & 0.39 & 0.02 & 0.59\\
		$\mu_{0,r}$ (mag arcsec$^{-2}$) & 1368 & -0.52 & -0.18 & 5.27 & 0.41 & -0.04 & 2.9 $\times$ 10$^{-6}$\\
		log(V$_{\text{last}}$ / km s$^{-1}$) & 1368 & 0.47 & 0.97 & -0.48 & 0.38 & 0.15 & 0.04\\
		log(V$_{\text{23.5}}$ / km s$^{-1}$) & 1368 & 0.43 & 0.65 & 0.32 & 0.40 & 0.09 & 0.05\\
		log($\Sigma^\star_{1}$ / M$_{\odot}$ kpc$^{-2}$) & 1368 & 0.44 & 0.27 & -0.64 & 0.39 & 0.01 & 0.78\\
		\hline
	\end{tabular}
\end{table*}

Based on the r$_P$ values, the central baryon-dominated structural parameter, $\mu_{0,r}$, is sightly more correlated with log($\mathcal{S}$) than the mostly DM-dominated parameter, V$_{\text{last}}$.
Although, none of these (log-log) correlations are especially strong, one would be hard pressed to single out a tightest distribution since the largest correlations are all within errors.
Some well-known trends stand out: low central surface brightnesses (mass densities) have shallow RCs (low $\mathcal{S}$) and (redder) galaxies with higher masses have higher RC slopes (\sec{Intro}).
That is, galaxies with small $\mathcal{S}$ have low stellar mass density, low concentration, low mass, and generally bluer colour relative to their higher $\mathcal{S}$ counterparts. 

In order to make better sense of the diversity in $\mathcal{S}$ from these putative correlations, 
we have conducted a Breusch-Pagan (hereafter BP) test for heteroskedasticity of each investigated structural parameter \citep{BPtest}. 
The BP test identifies a statistically significant dependence in the residuals of each relation as a function of the x-axis parameter. 
In essence, the BP test indicates whether diversity is dependent on a structural parameter of interest. 
The BP test p-values and slopes are given in \Table{Paramcorrelations}. 
Statistically significant variations in the scatters of log($\mathcal{S}$) along the x-axis are found for log(M$_{\text{tot}}$), $\mu_{0,r}$, log(V$_{\text{last}}$), and log(V$_{\text{23.5}}$). 
Additionally, the direction of scatter increase or decrease, given by the ``BP slope'', is calculated via a maximum likelihood estimation fit, assuming a linear trend in the scatter along the x-axis and normally distributed errors (as is also the case for the BP p-value).
Ultimately, the BP slope indicates the direction of scatter dependence along the abscissa of the relation, and the rate at which this scatter changes. 
For a structural parameter to have genuine diversity, statistically significant changes in scatter require a p-value~$< 0.05$. 

\Table{Paramcorrelations} also includes measures of intrinsic scatter for relations with log($\mathcal{S}$).
For more detailed comparisons between simulations and observations and sound conclusions regarding the ability of NIHAO and NIHAO-AGN simulations to reproduce the diversity of observed $\mathcal{S}$ and related scaling relations, we require all calculated intrinsic scatters to account for observational errors. 
These comparisons will be made in \sec{simulated_RCs}. 
While simulated galaxies have no observational errors, the intrinsic scatters inferred from our observed data must account for observational uncertainties. 
A basic accounting of all sources of uncertainty (adding in quadrature) for the relations of \Table{Paramcorrelations} suggests that those errors are not negligible. 
Further, there is considerable covariance between log($\mathcal{S}$) and velocity structural parameter measurements.
Thus, we employ a Bayesian intrinsic scatter estimation \citep{2021ApJ...912...41S} that accounts for all known sources of error and their full propagation through our analysis pipeline. 
This process involves resampling a dataset many times, perturbing each profile (RCs, light profiles, ellipticites, etc.) and parameter value (distance, intrinsic disk thickness, etc.) according to their associated errors, assuming they are normally distributed. 
The set of resampled datasets is then run through our analysis pipeline, allowing us to extract perturbed structural parameters which may then be included in a scaling relation and fit. 
The fit residuals are then used to create a probability density function of the Bayesian intrinsic scatter, from which $\sigma_i$ is taken to be the mode.
This yields a Bayesian intrinsic scatter, $\sigma_i$, for each of our relations. 
We neglect a classical intrinsic scatter analysis, as any source of error that is shared between the two axes (inclination, redshift, RC fitting, etc.) will be duplicated by the classical error propagation, unlike our Bayesian intrinsic scatter accounting.
For this reason, classical intrinsic scatter estimates ($\sigma_c^2 = \sigma_o^2 - \sigma_u^2$, where $\sigma_u$ is the quadratic sum of uncertainties) are biased low ~\citep{2021ApJ...912...41S}. 
The Bayesian intrinsic scatters of our observed relations are reported in \Table{Paramcorrelations} as $\sigma_i$. 

In general, we find the Bayesian intrinsic scatters of relations in \Table{Paramcorrelations} to be nearly identical to the associated total orthogonal scatters, differing by at most $\Delta\sigma = |\sigma_o - \sigma_i| = 0.02$ dex. 
This is because the scatter in the residual is mainly in the y-axis, and the error of the y-axis parameter, log($\mathcal{S}$), is relatively small. 
Thus, most points are scattered by observational errors along the relation (i.e., primarily along the x-axis). 
This results in an intrinsic scatter that is close to the total orthogonal scatter, and indicates that these relations are relatively robust to errors. 
We conclude that the total orthogonal scatter values of our observational data in parameter spaces with $\mathcal{S}$ are relatively observational error free, allowing for comparisons with simulations. 
The only exceptions to this are the parameters M$_{\text{gas}}$, M$_{\text{bar}}$, and f$_{\text{gas}}$, all of which contain a smaller sample size limited by the number of galaxies for which gas mass profiles are available. 
We therefore consider trends with these particular parameters with caution. 

The Bayesian intrinsic scatter is reported throughout the paper, in lieu of the total scatter. 
Our test results are reported in \Table{Paramcorrelations}. 
Three measures of correlation and diversity are now available: 
the Pearson's correlation coefficient, r$_P$; the total orthogonal scatter, $\sigma_o$ (not presented); and the intrinsic scatter, $\sigma_i$.  
These tell us about the tightness, and thus driver, of a relation, while the BP slope and p-value inform us about the heteroskedasticity (or change in scatter / amount of diversity) in log($\mathcal{S}$) along the relation. 
By virtue of the first three measures, log($\mathcal{S}$) correlates best with central surface brightness, $\mu_{0,r}$, 
and its BP p-value of 2.9$\times$10$^{-6}$ indicates that the scatter in $\mathcal{S}$ changes in a statistically significant manner along the relation. 
V$_{\text{last}}$ also displays a relatively tight relation with log($\mathcal{S}$) (r$_P = 0.47$), and a statistically significant diversity fluctuations given the BP p-value value of 0.04. 
Once again, one would be hard pressed to identify which of the baryon-dominated 
($\mu_{0,r}$) or DM-dominated (V$_{\text{last}}$) parameter correlates more tightly with, and thus drives, the diversity parameter $\mathcal{S}$.  

Our results can also be compared with \citet{kauff15}, who studied various drivers of inner rotation curve slopes, though our data probe a lower mass regime with a greater fraction of dwarf galaxies in our sample. 
Some of our findings are slightly at odds with \citet{kauff15} who found no correlation between their measures of inner RC slope and stellar mass, concentration, or stellar mass density. 
We propose that our differences stem largely from a broader parameter coverage: PROBES and PROBES-II galaxies span 5 orders of magnitude in stellar mass, while the \cite{kauff15} sample spans just 1.5 orders of magnitude in stellar mass. 
\cite{kauff15} also suggested that inner slopes are well correlated with galaxy age, a claim that cannot be verified with PROBES and PROBES-II galaxies short of obtaining SFR and $<$A$>$ parameters for them.
However, we have tested this claim with the NIHAO simulated data in \sec{simulated_RCs}.

\begin{figure}
	\includegraphics[width=\columnwidth]{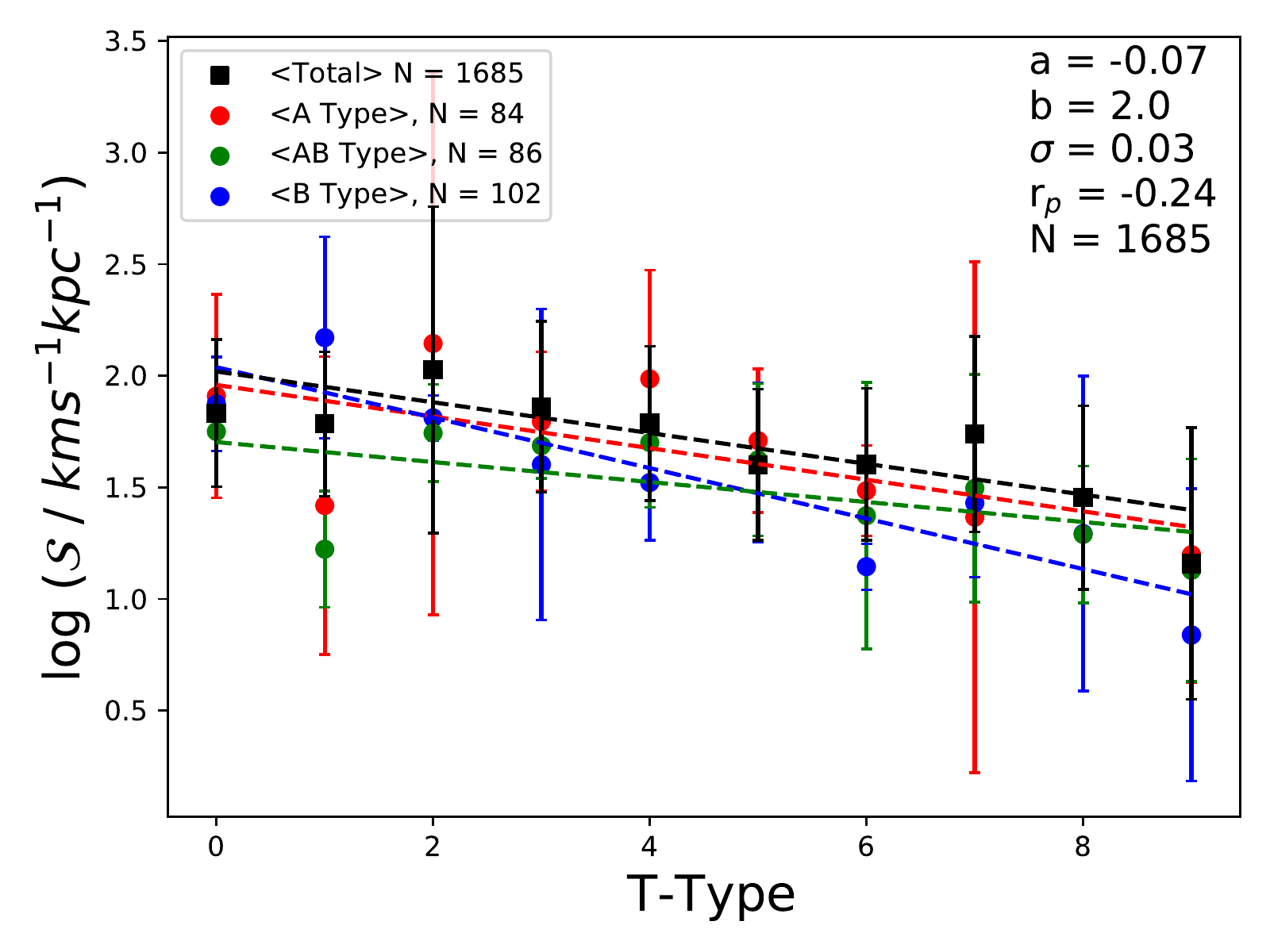}
    \caption{T-type against log($\mathcal{S}$) for all observed galaxies.
    Black points represent the median value of each Hubble T type value, with bars showing the interquartile range (IQR). 
    Red points display the median and IQR for barless galaxies, green points show intermediately barred galaxies, and blue points present barred galaxies. 
    Dashed lines show corresponding Orthogonal Distance Regression (ODR) linear fits. 
    The black, red, and blue dashed lines show distinct downward trends with increasing Hubble T type for late type galaxies, while the green dashed line shows no trend for intermediately barred galaxies. 
    The slopes of these relations are -0.05, -0.06, -0.03, and -0.01 for the black, red, blue, and green lines respectively. 
    Fit parameters for the total relation are shown in the top right corner of the figure.
    The legend displays the total number of galaxies in each fit. 
    Since not all galaxies were classified as barred / unbarred, the total N is larger than the sum of the components.}
	\label{fig:slope_Ttype}
\end{figure}

We now examine in \Fig{slope_Ttype} the dependence of $\mathcal{S}$ on morphology for late-type galaxies. 
The average $\mathcal{S}$ decreases with increasing T-type, in agreement with previous findings \citep{e16}, though the combination of PROBES and PROBES-II benefits from a larger sample.
No such correlation was found for early-type galaxies, though the sampling of ETGs in PROBES and PROBES-II is small and their RCs would require a more complete treatment than merely tracing emission lines \citep[see, e.g.,][]{o17}.
The log($\mathcal{S}$) - morphology relation of the observed LTG sample has a slope of -0.07, an intercept of 2.0, and a Pearson R correlation coefficient of -0.24. 
This suggests that galaxies with tightly wound spiral arms and larger bulges have higher $\mathcal{S}$ values, and baryons play a significant role in driving $\mathcal{S}$. 
PROBES and PROBES-II galaxies (see \sec{simulated_RCs}) follow the same general trend of decreasing $\mathcal{S}$ values with larger Hubble T type values. 
\Fig{slope_Ttype} also makes clear, across nearly all morphologies, that barless spirals (A type galaxies) have higher $\mathcal{S}$ values than barred spirals (B type galaxies) and intermediate AB type galaxies.   
Intermediate galaxies exhibit less change in log($\mathcal{S}$) with morphology, though small number statistics thwart a robust result.
The ``catch-all'' Hubble T-type 10 is neglected as it contains a variety of galaxy morphologies (mainly Irregulars). 

\cite{Sofue2016} suggested that a trend between barredness and inner RC slope may be due to the incorrect dynamical treatment of bars which biases rotational velocities low. 
Gas streaming along the bar may cause observed velocities to appear to be in solid body motion, yielding velocities that are lower than the true circular value \citep{buta1999,kuno2000,Sofue2016}. 
Furthermore bars may leave a wake as they rotate, slowing the bar itself and transfering angular momentum to the DM content in the center of the galaxy. 
This in turn may ``puff up'' the DM distribution, reducing the central dark matter fraction \citep{db10}. 
Observed velocities would thus be depressed in the inner regions where  $\mathcal{S}$ is measured. 
The statistical likelihood of viewing a bar side-on as opposed to end-on exacerbates this problem, suggesting that barred galaxies have systematically underestimated rotational velocities \citep{Sofue2016}. 
This is thus a measurement of the bias introduced in RCs via bars. 
This bias can only be remedied with more advanced RC extraction techniques \citep{spek07,2021MNRAS.502.3843S,Bisaria2021}. 
The small number of identified barred galaxies in the PROBES and PROBES-II sampled calls for caution when interpreting the effect of bars on RC shapes. 
Ultimately, the differences between the barred and unbarred galaxies are relatively weak, and larger galaxy samples would be needed to confirm any putative trends.  

We interpret all the findings above as evidence that $\mathcal{S}$ is a complex parameter, and that an identification of the main drivers of diversity in RC shapes based on our larger observational dataset alone remains challenging. 
Both baryon- and DM-dominated parameters appear to play significant roles in RC shape diversity, though tantalizing hints suggest that the baryonic content may be more dominant.  
We must turn to numerical simulations, which have no observational errors, in order to more closely examine the drivers of $\mathcal{S}$ and its diversity trends. 
We can also ask whether state-of-the art hydrodynamical simulations can replicate the diversity and general trends of the observed relations with $\mathcal{S}$ presented in \Table{Paramcorrelations}.

\section{Comparisons with the NIHAO Simulations}\label{sec:simulated_RCs}

We now analyze scaling relations between log($\mathcal{S}$) and various structural parameters, paying attention to any differences between the observed and simulated datasets. 
We wish to determine whether the simulated NIHAO galaxies reproduce the extent of log($\mathcal{S}$) found in the observed data, as well as the slope, intercept, scatter, and diversity of the observed scaling relations with log($\mathcal{S}$). 
This includes determining if NIHAO galaxies display comparable variations in velocity (or stellar mass) profile shapes relative to observed datasets. 
As shown in \cite{santos18}, the NIHAO simulations closely reproduce the observed V$_{\text{2kpc}}$ - V$_{\text{last}}$ plane, a common parameter space in which to investigate RC diversity. 
It is noted that this agreement was based on a more limited data set than we utilize here. 
We reproduce this test below (\sec{InnerSlopesAll}) and explore other parameter spaces related to inner rotation curve slopes, directed by our findings in \sec{Results}.  
Different relations will present different scatters and diversity signatures, and one must determine which simulated parameter spaces struggle to replicate observed data. 
Such relations can guide future refinement of subgrid physics for a more accurate approximation of nature.

We shall first proceed with an investigation of the effect of AGN on the shape of galaxy RCs, hinted at in \sec{agnslope}.
With the RCs of the 18 galaxies in common between NIHAO and NIHAO-AGN,
we can single out how AGN feedback influences $\mathcal{S}$ and to what extent it introduces diversity into galaxy RCs at fixed mass.

\subsection{Effect of SMBH feedback}
\begin{figure*}
    \includegraphics[width=\textwidth]{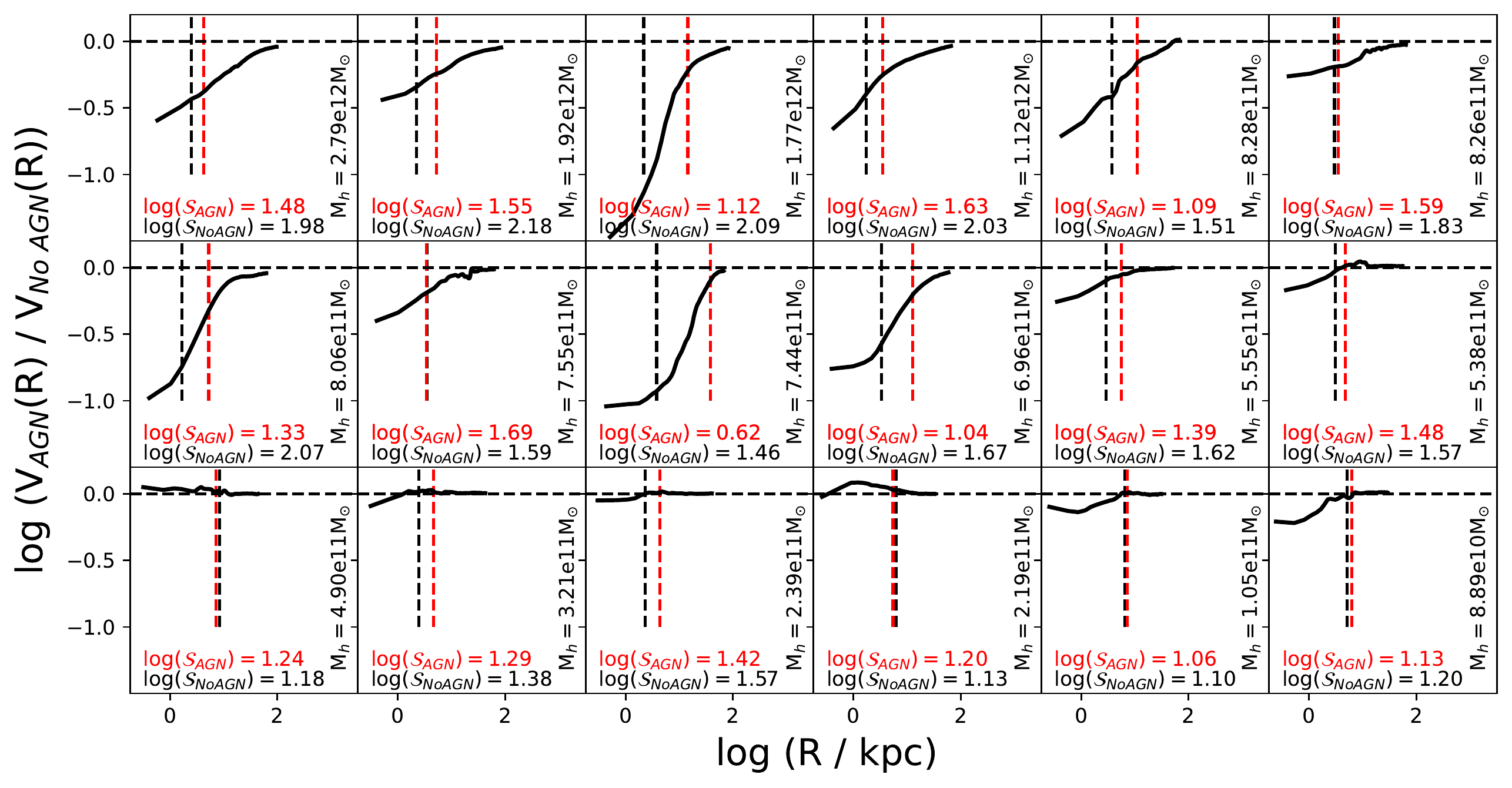}
    \caption{log(V$_{\text{AGN}}$(R) / V$_{\text{No AGN}}$(R)) for the 18 galaxies in common between NIHAO and NIHAO AGN. 
    Vertical black and red dashed lines show the PW transition radii for NIHAO and NIHAO-AGN fits, respectively. 
    The horizontal black dashed line shows the case V$_{\text{AGN}}(R) = $ V$_{\text{No AGN}}(R)$. 
    The measured inner slope, $\mathcal{S}$, is presented for both RCs; NIHAO and NIHAO-AGN values are shown in black and red text, respectively. 
    $\mathcal{S}$ is typically lower for NIHAO-AGN galaxies. 
    }
	\label{fig:NIHAO-NIHAOAGN}
\end{figure*}

In order to better understand the effect of SMBH feedback in shaping simulated RCs, we compare in \Fig{NIHAO-NIHAOAGN} the RCs for 18 (disc and irregular) galaxies in common with the NIHAO and NIHAO-AGN simulated datasets. 
Halos with a halo mass less than 5$\times$10$^{10}$ were not seeded with SMBH; any differences in the log($\mathcal{S}$) would therefore stem from random effects in the simulations. 
The 18 NIHAO-AGN galaxies presented here all have SMBHs. 

Ultimately, RC shapes vary most strongly for AGN hosting galaxies inhabiting more massive halos; for galaxies with small haloes, SMBH feedback has a lesser impact on RCs. 
AGN feedback lowers the gas mass, stellar mass \citep[also reported in][with FABLE]{2021MNRAS.503.3568K} and stellar mass density profiles of galaxies, in addition to significantly reducing rotational velocities in the largest halos. 
Galaxies with the strongest deviations between NIHAO and NIHAO-AGN typically display the largest fractional decrease in central baryonic mass. 
In all cases, the dark matter mass profile is the most constant between pairs, and gas mass profiles experience the largest deviations. 
The NIHAO-AGN simulations always produce mass profiles that are lower at all radii (within R$_{200}$) than their NIHAO counterparts, as noted by \citet{blank2019}. 
These differences are naturally most pronounced in the center of galaxies, within R$_{pw}$, but are still noticeable throughout the entire profile. 
SMBH feedback ultimately removes baryons from the central parts of a galaxy and redistributes DM, thereby reducing its central density \citep[see also][]{Maccio2020}. 
Thus, as expected, the log($\mathcal{S}$) values of NIHAO-AGN galaxies are smaller than the corresponding values of NIHAO simulations, averaging a decrement of 0.3 dex. 
This is particularly true for galaxies with a stellar mass of log(M$^{\star}/$M$_{\odot}) > 10$, beyond which $\mathcal{S}$ \textit{drops} substantially in NIHAO-AGN galaxies relative to NIHAO. 
The addition of SMBH feedback in the NIHAO-AGN simulations also significantly reduces the inner dark matter fraction by $\sim$20 percent and lowers the maximum velocity by $\sim$85 \kms on average (primarily by removing large bulge components). 
The last measured velocity point, here V$_{200}$, remains largely constant, but decreases by $\sim$11\kms on average. 
Ultimately, simulations reveal that an AGN can strongly alter a galaxy's RC, particularly within the central parts, increasing R$_{pw}$ by a factor of 2.5. 

The consistent drop in log($\mathcal{S}$) of NIHAO-AGN at fixed mass is incommensurate with observations (\sec{agnslope}) which suggest that RCs would be unaffected by AGN feedback. 
Indeed, observed AGN-hosting PROBES and PROBES-II galaxies show a broad range of log($\mathcal{S}$) with no consistent trend in RC shapes at fixed stellar mass (\Fig{AGN}). 

At this stage, we appreciate the scant nature of both observed and simulated data sets of AGNs with spatially-resolved RCs, and a resolution of this dichotomy most likely starts with significantly augmented data sets and a detailed understanding of all errors attributed to these galaxies.

\begin{figure}
    \includegraphics[width=\columnwidth]{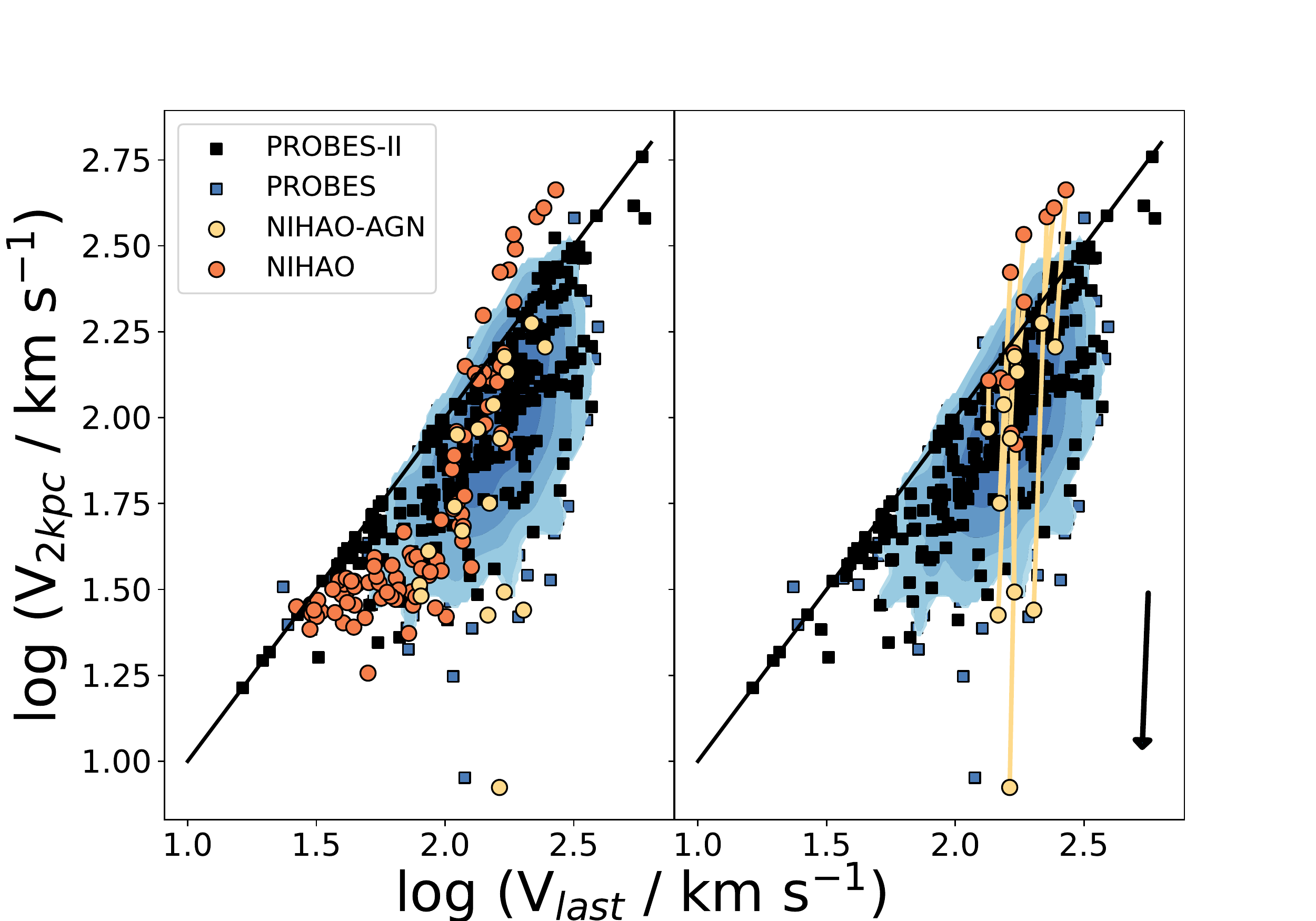}
    \caption{\textit{Left panel:} V$_{\text{2kpc}}$ - V$_{\text{last}}$ relation for the PROBES (blue contours and squares when outside the maximum extent of our Gaussian smoothed contours), PROBES-II (black squares), NIHAO (orange circles), and NIHAO-AGN (gold circles) galaxies. 
    Velocities for both observational and simulated RCs were estimated via the MP model. 
    NIHAO-AGN galaxies are only available for $\log V_{\text{last}} > 1.8$.
    The solid black line shows a one-to-one relation.
    \textit{Right panel:} Same as left, but only simulated galaxies in common to NIHAO and NIHAO-AGN are shown; all pairs are connected with a yellow line. 
    The downward black arrow in the bottom right corner displays the average effect on V$_{\text{2kpc}}$ and V$_{\text{last}}$ of addding an AGN to the simulated galaxies.}
	\label{fig:NIHAOV2kpc}
\end{figure}

\subsection{Inner RC Slopes of PROBES, NIHAO, and NIHAO-AGN}\label{sec:InnerSlopesAll}

The PROBES and PROBES-II datasets were used to explore the topic of galaxy diversity against numerical simulations with the largest possible sample of galaxies with available RCs and a larger stellar mass range. 
Specifically, we have tested if the NIHAO simulations can replicate the trends and distributions of various observed structural parameters (\sec{Results}). 

Based on comparisons with a smaller dataset, the NIHAO simulations were reported to generally reproduce the observed diversity of dwarf galaxy's RC shapes \citep{santos18}. 
These authors noted that NIHAO RCs reproduce the general trends of the observed V$_{\text{2kpc}}$ - V$_{\text{last}}$ relation, as well as the large observed scatter of low mass galaxies. 
For completeness, we have reproduced the V$_{\text{2kpc}}$ - V$_{\text{last}}$ plane for the NIHAO simulations with all PROBES and PROBES-II galaxies (regardless of available photometry) in \Fig{NIHAOV2kpc}. 
Contrary to \citet{santos18}, we found significant scatter in the observed V$_{\text{2kpc}}$ - V$_{\text{last}}$ plane across the entire mass range (not only at low masses), due in part to our larger observational sample.  
Furthermore, NIHAO simulated galaxies poorly trace the scatter of the observed data, especially at high mass where scatter is substantial.
This impression may partly result from the small number of high mass NIHAO galaxies relative to our observational samples. 
Importantly, the additional AGN feedback in NIHAO-AGN seems to push NIHAO galaxies onto the observed range, and better replicates the spread of the observed measurements.
This is most clearly seen in the right-hand panel of \Fig{NIHAOV2kpc}.
In particular, the addition of AGN feedback brings back NIHAO galaxies with V$_{\text{2kpc}}$ $>$ V$_{\text{last}}$) into the observed region of parameter space.
This suggests that AGN feedback is crucial at reproducing components of galaxy RC diversity. 

Linear fits of the combined observational data (PROBES and PROBES-II) in \Fig{NIHAOV2kpc} yield an orthogonal observed scatter of $\sigma_o = 0.13$ 
and a intrinsic scatter of $\sigma_i = 0.12$. 
Similar fits for the simulated NIHAO and NIHAO-AGN galaxies, which inherently display intrinsic scatter free of observational uncertainties, yield $\sigma_o = \sigma_i = 0.12$ (NIHAO) 
and $0.13$ (NIHAO-AGN), respectively.  
The observed and NIHAO intrinsic scatter values calculated assuming a linear relation are nearly identical along the V$_{\text{2kpc}}$ - V$_{\text{last}}$ relation, suggesting that observational errors can bias interpretations \citep{santos18}. 

However, NIHAO galaxies do not follow a linear relation; the curved trend of NIHAO galaxies is not matched by our observational data (whether detrended for observational errors or not), especially at the high mass end where strong departures are observed (i.e., where NIHAO galaxies have V$_{\text{2kpc}}$ $>$ V$_{\text{last}}$). 
This is determined by fitting a polynomial function to the NIHAO data, and calculating the 1$\sigma$ scatter in the residuals. 
The curved NIHAO trend encompasses at most $\sim 31$\% of the observational data (accounting for observational errors). 
All other observational data would require larger error values along both the x- and y-axes (by a factor of 3.2$\times$) to be placed within 1$\sigma$ of the NIHAO trend. 
Therefore, simulated galaxies do not closely replicate the observed V$_{\text{2kpc}}$ - V$_{\text{last}}$ parameter space. 

It is also apparent that the addition of SMBH feedback places some NIHAO-AGN simulations within the region of strongest ``inner mass deficit'', as defined by \citetalias{Oman2015}. 
The location in the V$_{\text{2kpc}}$ - V$_{\text{last}}$ space of NIHAO-AGN galaxies with V$_{\text{2kpc}}\ll$ V$_{\text{last}}$ suggests that more baryons must have been removed from their central regions than are currently found in those galaxies \citepalias{Oman2015}. 
Because mass depletion is seen in NIHAO-AGN galaxies, and not in their corresponding NIHAO analogs, we surmise that baryon induced core formation, particularly with the addition of AGNs among other feedback mechanisms (see \sec{SimData}), is consistent with the RC shapes of observed galaxies. 
This again indicates that AGN feedback is an important actor in regulating the RC shapes of high mass galaxies. 
Indeed, NIHAO-AGN galaxies land both within the observed V$_{\text{2kpc}}$ - V$_{\text{last}}$ relation, whilst simultaneously matching some of the most egregious outliers. 

Finally, the V$_{\text{2kpc}}$ - V$_{\text{last}}$ test of diversity must be treated with care since measurements are compared at fixed physical radii (e.g., 2 kpc) which may sample different environments over the range of galaxies. 
Physical scale comparisons certainly complicate the interpretation of RC shapes across a broad range of galaxy masses.
For this reason, we also considered other relations as tracers of RC diversity. 
To achieve this, we used the PW fitted inner RC slope, $\mathcal{S}$, which obviates this problem and appears to be a more robust indicator of RC diversity and shape in general. 
Hereafter, we used the PW function to analyze the RC shapes of PROBES, PROBES-II, NIHAO and NIHAO-AGN galaxies in greater detail.
In particular, we wish to understand whether simulated RC shapes also trace the distributions, trends, and scatters of other observational parameters, rather than the scatter in V$_{\text{2kpc}}$ and V$_{\text{last}}$ alone. 

To this end, we have reproduced the relations presented in \Table{Paramcorrelations} with the combined NIHAO and NIHAO-AGN sample. 
We have amalgamated the two samples to limit the effect of small scale statistics in the NIHAO-AGN sample. 
Since the simulated data have no measurement error, the orthogonal scatter, $\sigma_o$, about the fitted relations is the intrinsic scatter. 
For consistency with the observed galaxies in \Table{Paramcorrelations}, the same  parameters are presented for the combined NIHAO and NIHAO-AGN galaxies in \Table{scatters}. 
These include the Pearson's R correlation coefficient, slope, intercept, orthogonal scatter (equivalent to the intrinsic scatter), and BP slope statistic and p-value.
Some observational parameters, such as colour values, cannot be reproduced as a result of NIHAO having one photometric band; in such cases, the data were reported as ``-''. 
In general, the NIHAO and NIHAO-AGN galaxies have tighter linear correlations (defined by larger r$_{P}$ and smaller $\sigma_o$), but roughly similar slopes and intercepts when compared to the observed relations. 

\begin{figure*}
	\includegraphics[width=\textwidth]{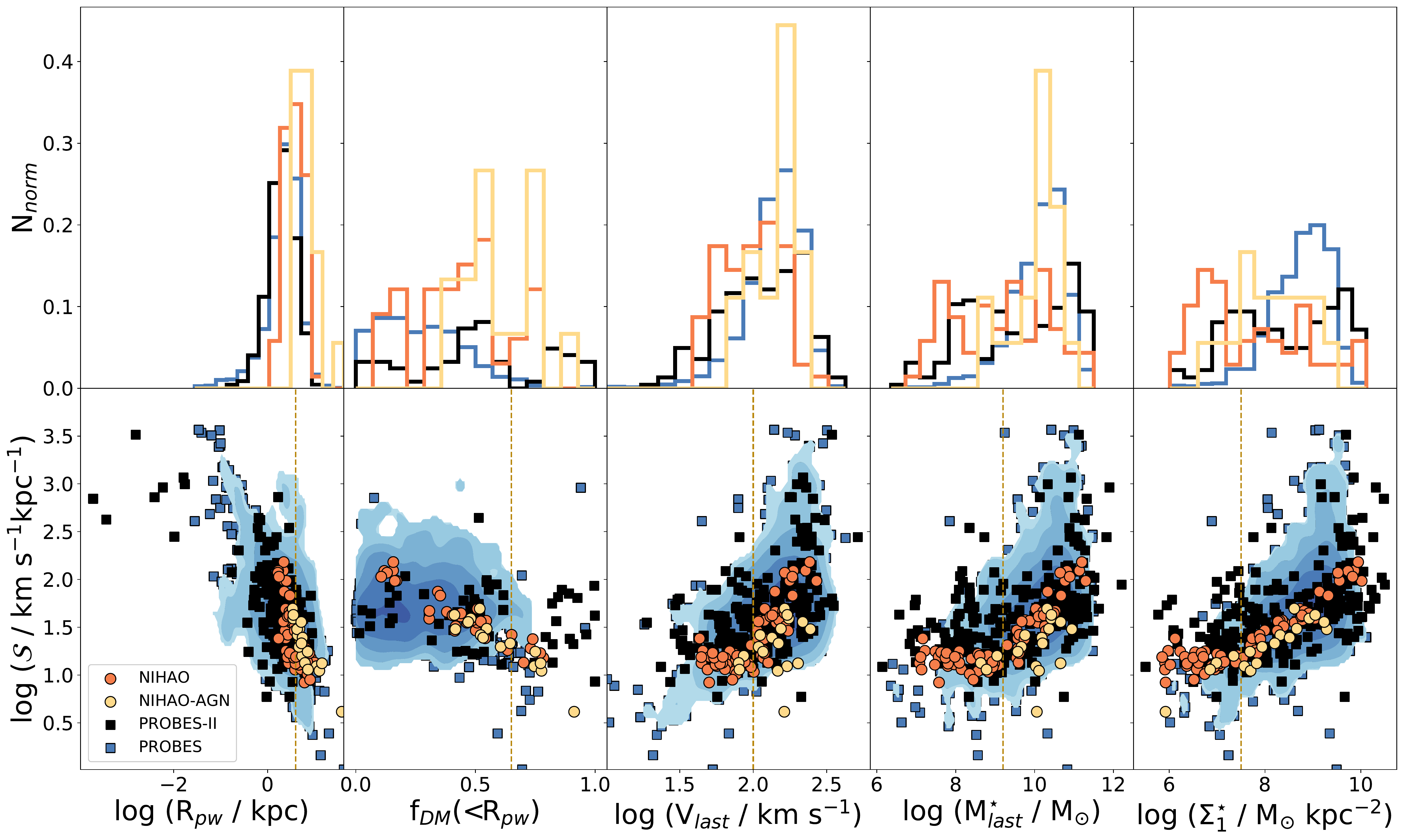}
    \caption{{\it Top row}: Normalized histogram distributions of structural parameters available in observed and simulated data. 
    {\it Bottom row}: log($\mathcal{S}$) plotted left to right against the piecewise (PW) fit transition radius, average DM fraction within the PW fit transition radius, last rotational velocity, total stellar mass, and stellar mass density within 1 kpc.
    Vertical dashed lines display the location of the turning points of NIHAO galaxies. 
    The average DM fraction within the PW fit transition radius is presented only for galaxies with log(M$^{\star}/M_{\odot}$) > 9.5, since gas masses are significant below this range.}
	\label{fig:NIHAOslopes}
\end{figure*}

Indeed, across all sampled scaling relations with log($\mathcal{S}$), the scatter values for the simulated data are lower than the observed scatters by roughly a factor of two, with the exception of $f_{\text{gas}}$, for which the observed sample size is smaller.
Since the observed intrinsic scatters are considered to be observational error free, we must conclude that this discrepancy is related to some limitations in the simulated data.
For an example of the differences between the simulated and observed datasets, \Fig{NIHAOslopes} shows log($\mathcal{S}$) for the PROBES, PROBES-II, NIHAO, and NIHAO-AGN samples plotted against various structural parameters. 
These parameters are: the piecewise transition radius, log(R$_{pw}$), the average dark matter fraction within R$_{pw}$, f$_{\text{DM}}(<R_{pw})$, the rotational velocity as measured by the MP model at the last measured radii, log(V$_{\text{last}}$), the stellar mass value at the last measured radii of the RC, log(M$^{\star}_{\text{last}}$), and the surface mass density as measured within 1 kpc, log($\Sigma^\star_1$), extending the analysis of \citet{santos20}. 
In most cases, NIHAO and NIHAO-AGN galaxies trace the observed distributions rather well, reproducing broad shapes and kinks in the observed data despite the smaller scatter than the observed data (see \Table{scatters}). 
This leads to the conclusion that NIHAO simulations struggle to match the diversity in these observed relations with $\mathcal{S}$. 

Close inspection of \Fig{NIHAOslopes} shows NIHAO failing to reproduce quickly rising compact massive galaxies with short turnover radii, as well as slowly rising low-mass galaxies, as noted in \cite{santos18}. 
For simulated galaxies to match the full range observed parameter distributions, one must occasionally invoke differences larger than observational errors. 
Based on estimated observational error values, a maximum of 82\% and an average of 71\% of observed galaxies fall within 1$\sigma$ of the NIHAO relation for each panel of \Fig{NIHAOslopes}.
However, the intrinsic scatters of the observed and simulated data are still not within error along any of these relations. 
Furthermore, roughly a third of observed galaxies are still found well outside of the relation defined by simulations, even when considering their observational errors, resulting in a significant mismatch. 
An incomplete treatment of observational errors can be excluded given the similarities between the observed $\sigma_o$ and $\sigma_i$, and thus a plausible solution to this mismatch involves a revised subgrid physics model in NIHAO which would broaden the relations, and induce greater scatter in log($\mathcal{S}$). 

\begin{table*}
	\centering
	\caption{Overview of all NIHAO structural parameters tested against $\mathcal{S}$ for diversity trends.
	Listed are the Pearson's correlation coefficient, r$_P$, the slope, $m$, the intercept, $b$, the orthogonal scatter, $\sigma_o$ ( = $\sigma_i$), the BP slope, and the p-values from the BP test. 
	Parameters which cannot be evaluated for the simulated galaxies are marked as ``-''. 
	} 
	\label{tab:scatters}
	\begin{tabular}{lrrrrrrc}
			\hline
		Simulated Parameter & {N \ } & {r$_P$ \ } & {m \ } & {b \ } & $\sigma_o$ & BP Slope & BP p-value\\
		\hline
		\hline
		log(M$_{\text{gas}}$ / M$_{\odot}$) & 87 & 0.49 & 0.21 & -0.74 & 0.27 & 0.07 & 0.04\\
		log(M$^{\star}_{\text{last}}$ / M$_{\odot}$) & 87 & 0.70 & 0.19 & -0.38 & 0.22 & 0.03 & 0.04\\
		log(M$_{\text{bar}}$ / M$_{\odot}$) & 87 & 0.60 & 0.23 & -0.95 & 0.24 & 0.05 & 0.04\\
		log(M$_{\text{DM}}$ / M$_{\odot}$) & 87 & 0.64 & 0.33 & -2.30 & 0.24 & 0.09 & 1.6 $\times$ 10$^{-3}$\\
		log(M$_{\text{tot}}$ / M$_{\odot}$) & 87 & 0.64 & 0.33 & -2.25 & 0.24 & 0.09 & 1.9 $\times$ 10$^{-3}$\\
		f$_{\text{gas}}$ & 87 & -0.29 & -2.81 & 1.53 & 0.29 & -1.99 & 0.11\\
		$f_{\text{DM}}(<R_{pw})$ & 87 & -0.91 & -1.18 & 2.13 & 0.13 & 0.05 & 0.11\\
		(g - z)$_{23.5}$ & - & - & - & - & - & - & -\\
		(g - z)$_{\text{1 kpc}}$ & - & - & - & - & - &- & -\\
		C$_{28}$ & - & - & - & - & - & - & -\\
		C$^\star_{28}$ & 69 & 0.73 & 0.18 & 0.66 & 0.22 & 0.06 & 8.6 $\times$ 10$^{-3}$\\
		$\mu_{0,V}$ (mag arcsec$^{-2}$) & 69 & -0.92 & -0.10 & 3.67 & 0.12 & 0.01 & 0.49\\
		log(V$_{\text{last}}$ / km s$^{-1}$) & 87 & 0.80 & 0.93 & -0.58 & 0.19 & 0.07 & 0.41\\
		log(V$_{\text{23.5}}$ / km s$^{-1}$) & 69 & 0.80 & 0.35 & 0.85 & 0.19 & 0.10 & 1.2 $\times$ 10$^{-4}$\\
		log($\Sigma^\star_{1}$ / M$_{\odot}$ kpc$^{-2}$) & 87 & 0.88 & 0.24 & -0.49 & 0.14 & -0.01 & 0.52\\
		\hline
	\end{tabular}
\end{table*}

Both the NIHAO and NIHAO-AGN galaxies exhibit distinct turning points in the relations presented in \Fig{NIHAOslopes} that are reflected (albeit more faintly) in the observational data. 
These are identified via PW fits, where the fitting function is essentially the same as \Eq{piecewiseRC}. 
The clearest turning point is seen in the plane of $\log($M$_{\text{last}}^{\star})$ vs. $\log$($\mathcal{S}$). 
Galaxies with log(M$^{\star}/$M$_{\odot}) \lesssim 9.3$ have nearly constant $\mathcal{S}$, before rising sharply for larger stellar masses. 
This turning point also corresponds to $V_{\rm max} \sim 100$ km~s$^{-1}$, a piecewise RC fit transition radius of about $4$ kpc, a value of $f_{\text{DM}}(<R_{pw}) \approx 0.65$, or $\Sigma^\star_1 \sim 10^{7.8} $M$_{\odot}/$kpc$^{-2}$.
Incidentally, as shown below, this turning point also corresponds to a total mass of $\log($M$_{\text{tot}}/$M$_{\odot}) \sim 10.8$, or roughly SFR $\sim$ 0.13~$M_{\odot}$ yr$^{-1}$ (for NIHAO galaxies) (\sec{sfr}).
A clustering analysis indicates that both the simulated and observed data are best described by at most two clusters, via the traditional ``elbow method'' \citep{Shi2021}. 
The transition from roughly constant to sharply increasing log($\mathcal{S}$) appears to scale firmly with the dark matter fraction of the galaxy. 
This observation calls for identifying the drivers of these turning points with the help of galaxy mass profiles (see \sec{MassProfs}). 

Following our analysis in \sec{Results}, we have performed a BP test on the various structural parameter projections with the simulated data in \Fig{NIHAOslopes} to determine the strength of predicted RC inner slope diversity.
For the combined NIHAO samples, we find a statistically significant change in scatter along the relations between $\mathcal{S}$ and all measures of galaxy mass, as well as C$^\star_{28}$ and log(V$_{23.5}$). 
The comparison of the diversity fluctuations between the simulations (\Table{scatters}) and observed data (\Table{Paramcorrelations}) also shows significant differences.
The BP p-values of the NIHAO and observed data only agree (in statistical significance) for five parameters: log(M$^{\star}_{23.5}$), $f_{\text{DM}}(<R_{pw})$, $f_{\text{gas}}$, C$^\star_{28}$, and log(V$_{23.5}$). 
In particular, the simulated galaxies predict significant diversity in scatter along all relations pertaining to galaxy mass, while the scatter of log($\mathcal{S}$) for observed galaxies scales only with stellar mass. 
Given the small variations in the BP slopes, little difference is found between the simulated and observed data; scatter for both samples typically increases gently (shallow slopes) as a function of the x-axis parameter. 
Contrary to this statement, the scatters in the f$_{\text{gas}}$ - log($\mathcal{S}$) relation for the observed and simulated datasets decrease sharply as a function of f$_{\text{gas}}$.
We believe this is due to primarily to the large percentage of low gas mass galaxies in the subsample of PROBES-II with gas masses. 
None of the relations presented in \Fig{NIHAOslopes} have strongly correlated residuals (all have r$_{P} < 0.25$) with all tested parameters. 

While NIHAO and NIHAO-AGN reproduce the general trends of many scaling relations found in the observed data, they often fail to reproduce the full diversity in RC shapes between spiral galaxies of similar maximum rotation velocity (or fixed $M_{\text{star}}$) in all parameter spaces, even when observational errors are taken into account. 
Even if observational errors are underestimated, particularly in RC profiles \citep{2021MNRAS.502.3843S}, one would require extreme excursions to match some observations with simulations. 
We surmise that the disparity between simulations and observations is due to the incomplete subgrid physics models in numerical (NIHAO) simulations. 

\subsection{Inner RC Slope, Star Formation Rates, and Age}\label{sec:sfr}

\begin{figure*}
	\includegraphics[width=\textwidth]{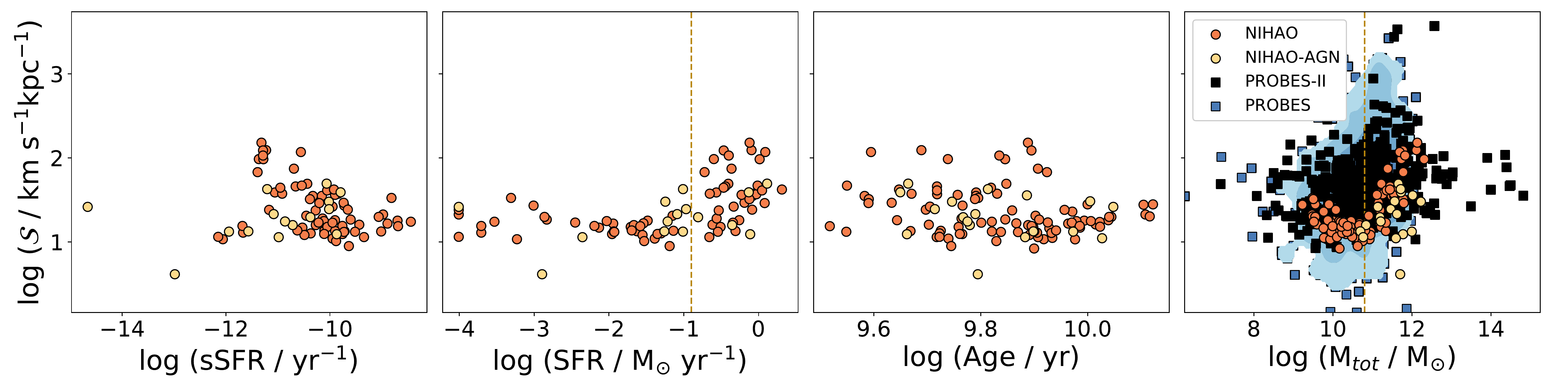}
	\caption{From left to right, inner RC slopes of NIHAO galaxies versus specific star formation rate (sSFR), star formation rate (SFR), and stellar age. 
	In the right-most plot, the inner RC slope is shown against total mass for PROBES (blue points), PROBES-II (red points), NIHAO (orange points), and NIHAO-AGN (gold points). 
	Vertical dashed lines display the location of the turning points of the NIHAO galaxies.}
	\label{fig:NIHAO_SFR}
\end{figure*}

NIHAO galaxies have also been shown to reproduce the observed star formation main-sequence \citep{arora2021}. 
The relation between star formation rate (SFR) and log($\mathcal{S}$) of NIHAO and NIHAO-AGN galaxies is thus also worth exploring. 
\citet{kauff15} suggested that measures of age correlate with inner RC rise, where galaxies with slowly rising inner RC slopes have younger central stellar populations. 
While age measurements are currently unavailable for our observational datasets, the reported correlation between central RC slope and stellar age by \citet{kauff15} is absent in the NIHAO simulations, as seen in \Fig{NIHAO_SFR}. 

However, an interesting relation arises between log($\mathcal{S}$) and sSFR, with higher scatter at higher sSFR. 
SF might boost the RC shape scatter at low mass: enhanced gas inflows and velocity dispersion \citep{Hung_2018} linked to star formation would disturb the velocity field. 
\cite{2016MNRAS.462.3628R} have also suggested that stellar feedback and starbursts could create large \HI~bubbles that distort the observed RC shape, contributing scatter to the log($\mathcal{S}$) and sSFR relation. 
With the NIHAO and NIHAO-AGN galaxies, we confirm that star formation processes correlate with high log($\mathcal{S}$) values, increased scatter, and diversity in low mass galaxies. 

NIHAO and NIHAO-AGN galaxies also exhibit turning points in the plane of log($\mathcal{S}$) and measures of star formation.
These turning points match the suggestion by \citet{2004ApJ...608..189D} that SF is less efficient below $V_{\text{circ}} < 120$ \kms. 
These authors proposed, based on the presence or absence of dust lanes in disk galaxies, that this threshold represents a drop in the gas scale height of the disk, which in turn leads to increased SFR when disk instabilities are present. 
Galaxies with rotational velocities below $120$ \kms\ are stable and lie below the Kennicutt SF density threshold. 
Given that the NIHAO simulations use the Kennicutt-Schmidt SF law as a constraint, similar thresholds are indeed expected.
The log($\mathcal{S}$) versus SFR relation confirms that galaxies with lower SFRs have lower log($\mathcal{S}$) values.
A turning point in this relation is found at roughly 0.13 $M_{\odot}$ yr$^{-1}$; all galaxies with inner slopes log($\mathcal{S}$)$ > 1.6$ lie beyond this SFR threshold. 
Incidentally, this also corresponds to a turning point in total mass, with the threshold at log(M$_{\rm tot}/$M$_{\odot} > 10.8$) (see \Fig{NIHAO_SFR}), reconfirming the trends found in \Fig{NIHAOslopes}. 
NIHAO-AGN galaxies are found to follow relations similar to NIHAO, though the SFR is on average reduced, suggesting that the inclusion of SMBH feedback provides a modest decrease of SFR in most galaxies. 
Recent related studies indicate little to no observed relation between SFR and AGN activity \citep[see][]{2012ApJ...760L..15H,2017MNRAS.472.2221S,2017ApJ...841..102S,2019MNRAS.486.4320R}. 
Quantitatively, we find that the average change between NIHAO and NIHAO-AGN is $\Delta$log(SFR) $ =-0.77$ and $\Delta$log(sSFR) $ =0.04$ (largely negligible). 
No trend is found for galaxy age; a negligible reduction in total mass ($\Delta$log(M$_{\text{tot}}$) $ =0.06$) is identified. 

\begin{figure}
	\includegraphics[width=\columnwidth]{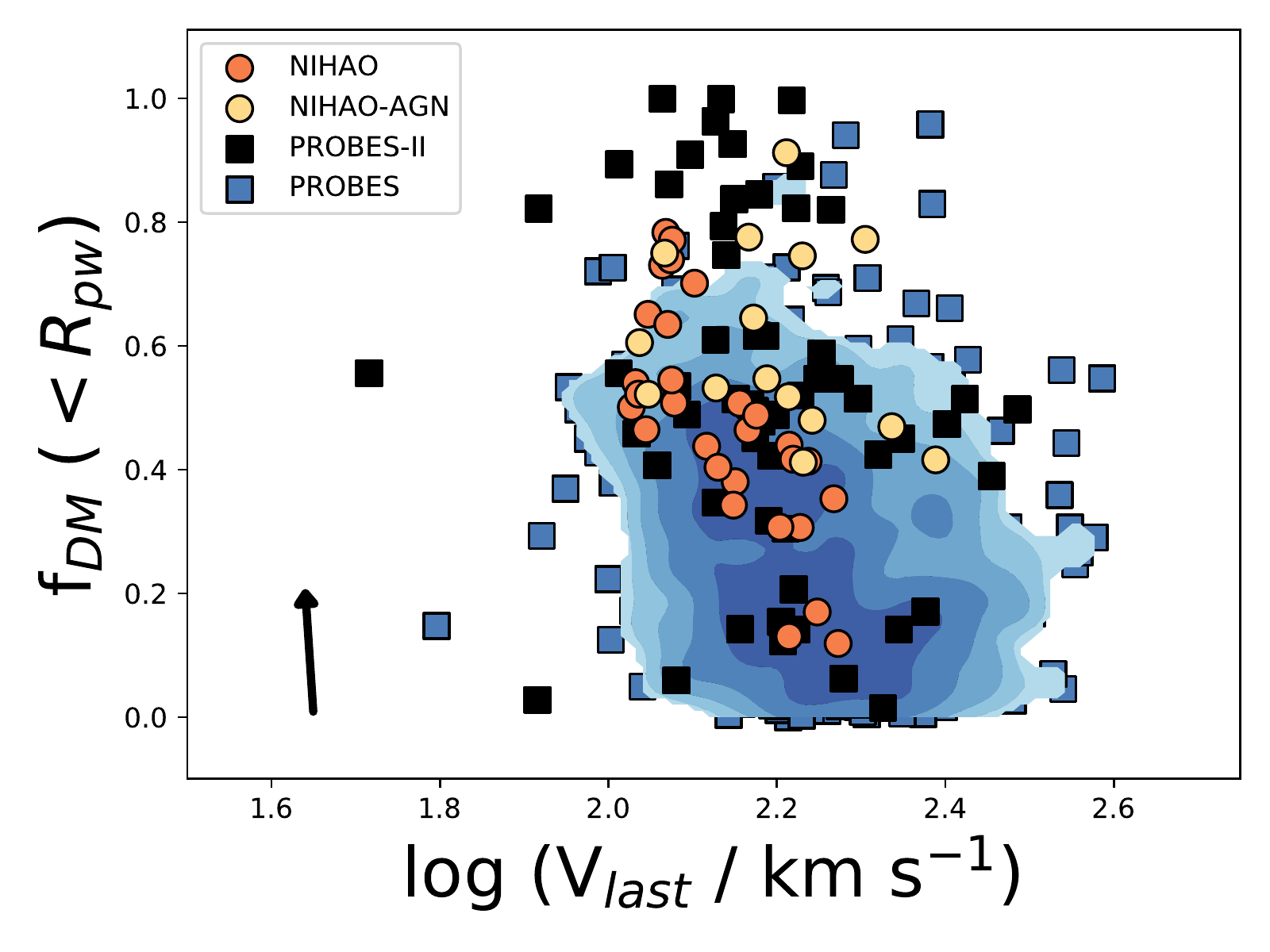}
    \caption{$f_{\text{DM}} (<R_{pw})$ against $V_{\text{last}}$. 
    Overall, NIHAO and NIHAO-AGN galaxies reproduce the properties of centrally DM-dominated galaxies fairly well, but are a poor match to the centrally baryon-dominated slow rotators. 
    Only galaxies with log(M$^{\star}/M_{\odot}$) > 9.5 are displayed, to ensure that the lack of gas mass profiles for some galaxies does not impact the final values of f$_{DM}$. 
    The black arrow shows the average displacement from NIHAO to NIHAO-AGN. } 
	\label{fig:COMPARE_ISfdm}
\end{figure}

\subsection{Inner RC Slope Baryon Dependence} 

We now wish to quantify the impact of the DM fraction on the simulated RC shapes. 
However, extensive observational errors in M$_{\text{tot}}$, sometimes as large as one dex, make the recovery of $f_{\text{DM}} (<R_{pw})$ challenging \citep[][]{Maccio2020}. 
The expected relationship between log($\mathcal{S}$) and the average DM fraction within the transition radius of the piecewise fit is cautiously explored in \Fig{NIHAOslopes}. 
On average, NIHAO galaxies with $\mathcal{S}$ of $10$ km s$^{-1}$ kpc$^{-1}$ have $\approx 80$\% more DM (relative to baryons) in their inner regions than galaxies with $\mathcal{S}$ of $100$ km s$^{-1}$ kpc$^{-1}$. 
This reminds us that higher (lower) inner slopes, as measured by $\mathcal{S}$, are predominantly found in galaxies that are more baryon (DM) dominated in the inner regions (see also the first paragraph of \sec{Intro}). 

Motivated by the use of the V$_{\text{2kpc}}$ - V$_{\text{last}}$ relation as a proxy for the inner RC slope, we also show the $f_{\text{DM}} (<R_{pw})$ - V$_{\text{last}}$ relation in \Fig{COMPARE_ISfdm}. 
The latter makes clear that the simulated galaxies struggle at reproducing the full observed scatters along both relations. 
Comparisons between the samples must account for observational errors, and we follow a similar analysis to that performed in \Table{scatters}, where $\sigma_o$ is calculated for the observational and simulated samples. 
While the observed error in $f_{\text{DM}} (<R_{pw})$ is admittedly large, the errors in $\mathcal{S}$ and log(V$_{\text{last}}$) are generally quite small. 
The intrinsic and total orthogonal scatters for the combined observed samples is $\sigma_o = 0.20$ and  
$\sigma_i = 0.13$ 
for and $f_{\text{DM}} (<R_{pw})$ - log(V$_{\text{last}}$). 
Similarly, for the combined simulated data, we find 
$\sigma_i = \sigma_o = 0.13$. 
Therefore, the intrinsic scatters are identical in the $f_{\text{DM}}~(<R_{pw})$ - log(V$_{\text{last}}$) plane. 
This highlights once again the ability for simulations to reproduce some parameter projections, while struggling at others. 

A few NIHAO-AGN galaxies are scattered significantly in the $f_{\text{DM}} (<R_{pw})$ - log(V$_{\text{last}}$) space relative to their the NIHAO counterparts.
The scatter is primarily in $f_{\text{DM}} (<R_{pw})$. 
This indicates that SMBH feedback can cause significant variations in $f_{\text{DM}} (<R_{pw})$ at fixed log(V$_{\text{last}}$), reducing the inner DM fraction. 
Recall that adding SMBH feedback to the NIHAO simulations significantly reduces the inner dark matter fraction by typically $\sim$20 percent, lowers $\mathcal{S}$ by $\sim$0.3 dex, and lowers the maximum velocity by $\sim$85 \kms. 
This suggests that log(V$_{\text{last}}$) (and $\mathcal{S}$) is primarily driven by the baryonic content (or lack thereof) within R$_{pw}$, as expected. 

The $f_{\text{DM}} (<R_{pw})$ - log(V$_{\text{last}}$) relation also presents a compelling trend for simulators and observers alike, particularly because the intrinsic scatters are calculated to be identical between our simulations and observations. 
This panel shows a significant diversity in rotational velocity (and therefore mass) in baryon-dominated systems that \textit{is} reproduced by the NIHAO simulations. 
However, NIHAO galaxies have few baryon-dominated slow rotators, likely due to the composition of the NIHAO and NIHAO-AGN simulations, which are dominated by lower mass, dark matter dominated galaxies. 
A challenge for future simulations is indeed to reproduce the RC shapes of slow rotators.

\section{Diversity of Galaxy Stellar Mass Profiles}\label{sec:MassProfs}

We have characterized above the diversity in RCs of both observed and simulated galaxies. 
However, it is anticipated that any contribution of baryons to the galaxy inner RC shapes should translate into some diversity in galaxy stellar mass profiles. 
Therefore, to enable better comparisons of the latter with simulations, we have used the stellar surface mass density measured within 1 kpc, $\Sigma^\star_1$ [M$_{\odot}$ kpc$^{-2}$],  as an analogue to the inner mass profile slope.  $\Sigma^\star_1$ can be easily extracted from observations and simulations alike. 

\begin{figure*}
 	\includegraphics[width=\textwidth]{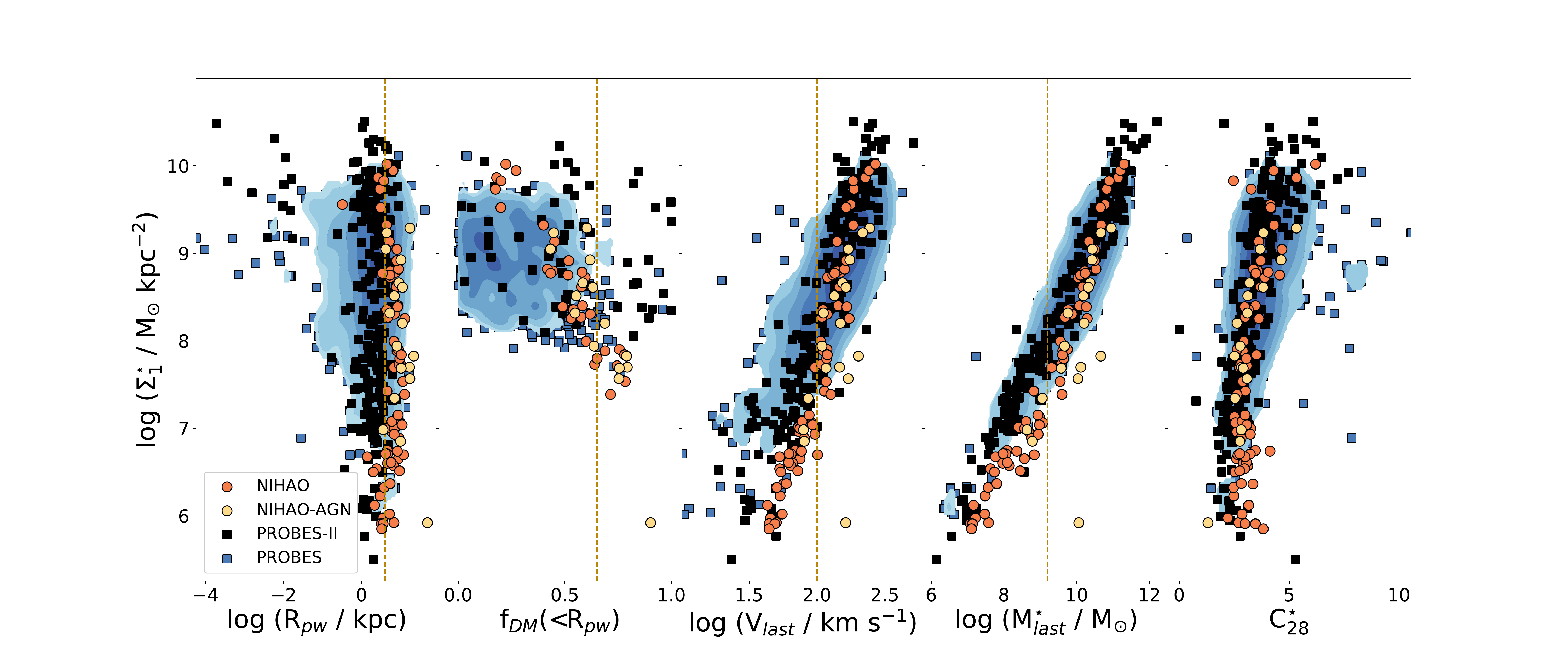}
    \caption{Stellar log($\Sigma^\star_1$) of PROBES-II, PROBES, NIHAO, and NIHAO-AGN galaxies formatted as in \Fig{NIHAOslopes}. 
    The vertical dashed line displays the turning points found in \Fig{NIHAOslopes}. 
    The average DM fraction within the piecewise fit transition radius is presented only for galaxies with log(M$^{\star}/M_{\odot}$) > 9.5, ensuring that the lack of gas mass profiles for some galaxies does not impact the final values of f$_{\text{DM}}$.
    } 
    \label{fig:Mslopes}
\end{figure*}

Various relations between log($\Sigma^\star_1$) and previously identified structural parameters of interest are presented in \Fig{Mslopes}. 
As before, PROBES, PROBES-II, NIHAO, and NIHAO-AGN are all included in this analysis. 
Here, log(R$_{pw}$) remains the piecewise transition radius as defined by the RC; in this case, little correlation is found with log($\Sigma^\star_1$). 
These relations are far tighter, with less intrinsic scatter in each projection, than those determined with log($\mathcal{S}$) in \Fig{NIHAOslopes}. 
Unlike log($\mathcal{S}$), it is comforting that the range of observed log($\Sigma^\star_1$) is well matched by the simulations.
However, the intrinsic scatters between simulations and observations are still a poor match for some of the projections examined. 
This is especially true for projections against $f_{\text{DM}} (<R_{pw})$; once again, the NIHAO and NIHAO-AGN simulations struggle to reproduce the range of observed $f_{\text{DM}}$. 

Unlike many observed trends with $\mathcal{S}$, and for all projections other than $f_{\text{DM}}$, the NIHAO + NIHAO-AGN relations replicate the slopes and intercepts of all observed projections within error.
The ``diversity problem'' is thus less conspicuous for scaling relations that depend on the inner stellar mass slope, rather than those relative to the inner RC slope.
Broadly speaking, the NIHAO + NIHAO-AGN simulations can more successfully reproduce baryon-dominated structural parameters and related scaling relations.
Poorer matches between simulations and observed data are obtained with DM-dependent quantities, such as $\mathcal{S}$ and $f_{\text{DM}} (<R_{pw})$.
Furthermore, while relations with stellar log($\Sigma^\star_1$) and log($\Sigma^\text{DM}_1$) remain relatively tight, we have verified that the same relations with log($\Sigma^{\text{gas}}_1$) display greater scatter, indicating that gas fractions may contribute significantly to galaxy profile shape diversity.

We also pay attention in \Fig{Mslopes} to the inflexion points found in the scaling relations with $\mathcal{S}$ (see \Fig{NIHAOslopes}); these points are represented by dashed gold lines. 
Recall that these turning points in log($\mathcal{S}$) were found to be mainly due to the increased dark matter fractions in lower mass galaxies (\sec{InnerSlopesAll}).  
However, these turning points are now largely absent in \Fig{Mslopes} for all projections against log($\Sigma^\star_1$).  

Close inspection of \Fig{Mslopes} shows simulations generally reproducing stellar mass profiles across the entire mass range. 
Based on estimated observational error values, a maximum of 99\% (in the log($\Sigma^\star_1$) - log(M$^{\star}_{\text{last}}$) projection) and an average of 78\% of observed galaxies fall within 1$\sigma$ of the simulated relations for each panel of \Fig{NIHAOslopes}.
However, generally, the observational data and the associated observational errors in these projections with log($\Sigma^\star_1$) are far closer to the trend reproduced by the simulated galaxies than in the projections with log($\mathcal{S}$).
Thus, we can conclude that NIHAO and NIHAO-AGN reproduce the stellar mass distribution relatively well. 

In conclusion, we have contrasted the relative inability of NIHAO to replicate the full range of observed galaxy RCs (\sec{simulated_RCs}), with NIHAO's success to more faithfully replicate the stellar mass profiles of observed galaxies, with little evidence of diversity. 
\section{Summary}\label{sec:Conclusions}

In an effort to characterize the diversity of spiral galaxy RCs across a wide range of stellar masses, a large sample of observed RCs and light profiles for 1752 galaxies (PROBES + PROBES-II) and numerical simulations of $\sim$100 galaxies from NIHAO and NIHAO-AGN were assembled. 
Motivated by \citetalias{Oman2015}, our study is a six-fold increase upon their original sample of 304 spiral galaxies and is supplemented with light profile information. 
We have improved upon previous studies of galaxy diversity by: 
(i) assembling a large sample of spiral galaxy RCs;
(ii) acquiring uniform, and deep, photometric data from the DESI and SDSS surveys; 
(iii) developing an extensive set of RC and light profile characterizations of galaxy diversity and profile shapes;
(iv) robustly comparing these characterizations of galaxy diversity with numerical simulations; as well as isolating the effect of an AGN in these numerical simulations; and 
(v), separating observed from intrinsic trends in the space of galaxy diversity following \citet{2021ApJ...912...41S}.

We have determined that galaxy inner RC shapes are more strongly correlated with baryon dominated structural parameters than with dark matter dominated parameters. 
RCs were modeled with a piecewise (PW) function in order to derive their inner and outer slopes, which were compared among stellar mass bins. 
Two distinct observed galaxy populations were found: 
(i) low mass galaxies of bluer $(g-z)$ colour with correspondingly low surface brightness (stellar mass density), concentrations and higher dark matter fractions, and 
(ii) higher mass galaxies of redder $(g-z)$ colour, high surface brightness (stellar mass density), higher concentrations, and lower dark matter fractions. 
These two populations straddle a stellar mass boundary of 
log(M$^\star/$M$_{\odot})\sim 9.3$, or equivalently log($\Sigma^\star_1 / $M$_{\odot}$ kpc$^{-2}$) $\sim 7.8$, a value of $f_{\text{DM}}(<R_{pw}) \approx 0.65$, 
a total mass of $\log($M$_{\text{tot}}/$M$_{\odot}) \sim 10.8$, and a circular velocity of log(V$_{\rm max}) \sim 2$. 
Additionally, we identify a transition at a SFR of roughly 0.13~$M_{\odot}$ yr$^{-1}$ with the NIHAO and NIHAO-AGN galaxies. 
Bluer galaxies follow a trend of slowly increasing log($\mathcal{S}$) with increasing stellar mass surface density, while redder galaxies show a steep RC rise matched by a rapid increase in stellar mass surface density. 
Across both populations, the diversity (as gauged by the BP slope) is roughly comparable, or mildly increasing, over all mass ranges in the observational data once observational errors are taken into account (see \Fig{RCbinneddiversity}, \Fig{NIHAOV2kpc}, \Fig{NIHAOslopes} and \Fig{VMR_all}). 
NIHAO and NIHAO-AGN faithfully reproduce these general trends and turning points in many of the observed relations. 

While selected differently and sampling slightly different galaxy populations, the PROBES and PROBES-II catalogues have similar diversity signatures (differing by at most 0.07~dex) when observational errors are taken into account. 
On the other hand, the simulated NIHAO and NIHAO-AGN galaxies occasionally have tighter intrinsic parameter distributions, and struggle to reproduce the diversity and intrinsic scatter of observational data. 
This is indeed the root of the diversity problem affecting numerical simulations of galaxies. 
Even with observational errors taken into account, simulated galaxies cannot reproduce the full diversity of RC shapes and DM-dominated parameters.
The inclusion of AGN feedback in the NIHAO-AGN simulation is insufficient to fully solve this predicament, though it serves as an improvement; i.e. placing galaxies on the V$_{\text{2kpc}}$ - V$_{\text{last}}$ relation closer to the observations. 
The differences between the (intrinsic) observational diversity and the one seen in the simulations is thus most likely due to shortcomings in the simulation's subgrid physics. 

AGN hosts may occasionally disrupt kinematics in higher mass galaxies, and reduce $\mathcal{S}$ values as found in the NIHAO-AGN simulations relative to NIHAO counterparts by $\sim$0.3 dex. 
The addition of SMBH feedback in the NIHAO-AGN simulations also significantly reduces the inner dark matter fraction by $\sim$20 percent and lowers the maximum velocity by $\sim$85 \kms. 
The inclusion of AGN in simulations confirms their significant impact on galaxy RC profiles, particularly at halo masses greater than 5$\times$10$^{11}$M$_{\odot}$. 
While the tests presented here are tantalizing, more work is needed to fully characterize the effects of SMBH feedback on gas kinematics in galaxies.
AGN searches should be conducted in low mass galaxies in order to quantify their impact on galaxy velocity fields, as some observations suggest \citep{2019ApJ...884...54M, Manzano_King_2020,2020ApJ...898L..30M}. 

Finally, we investigated the stellar mass profiles of both simulated and observed galaxies with the use of $\Sigma^\star_1$. 
It is found that simulated galaxies follow the relations described by the observed population more tightly in the $\Sigma^\star_1$ parameter projection than for RC profiles shapes. 
The lack of diversity in the stellar mass profiles (and dark matter mass profiles) indicates that the gas matter plays a pivotal role in the variation of log($\mathcal{S}$). 
Additionally, given that the relations with $\Sigma^\star_1$ and $\Sigma^{\text{gas}}_1$ do not show the turning points found in relations with log($\mathcal{S}$), but that $\Sigma^{\text{DM}}_1$ does, we surmise that the turning points in log($\mathcal{S}$) are mostly driven by changes in the dark matter fractions of lower mass galaxies. 
Ultimately, the comparisons between observational and simulated RCs and stellar mass profiles provides useful indication to observers and theorists alike for the role of stellar, gas, and dark matter components in the RC diversity problem. 

While we have characterized the differences between NIHAO simulations and the PROBES and PROBES-II observations, future work must further establish whether the inability of NIHAO and NIHAO-AGN (or any other galaxy simulations) to reproduce the range of RC and light profile shapes at fixed mass is related to:
a) the actual backbone of galaxy formation: $\Lambda$CDM + hierarchical assembly; and/or 
b) limitations of subgrid model baryonic physics.

Option (a), alluded to in \citetalias{Oman2015}, posits that CDM produces self-similar DM density profiles, and therefore points towards the self-similarity of simulated galaxy RCs.
If all other options are excluded, this could represent a great challenge to our understanding of galaxy formation. 
Option (b), would require improved models of baryonic physics, such as those critical for shaping galaxy RCs, such as such as star formation, BH accretion, and AGN feedback, among others. 
NIHAO does create galaxies with some RC diversity \citep{santos18,santos20},
but the requirement to match both simulations and observations for various structural parameter projections involving the inner RC slope is more complicated than previously stated or envisioned. 
NIHAO-AGN, with the addition of AGN feedback, is a step in the right direction, and future work should include improved and expanded feedback mechanisms. 

Therefore, while the CDM backbone is solid, the modeling of the interplay between the dark matter and baryonic components in simulations needs improvements. 
Given that galaxies are non-linear systems, solving for option (b) is not straightforward, but observations and numerical tests (e.g., AGN implementation) like the ones reported here will guide future developments. 

\section*{Acknowledgements}\label{sec:acknow}

We are grateful to Queen’s University and the Natural Sciences and Engineering Research Council of Canada for critical support through various scholarships and grants. 
Special thanks to Kyle Oman for providing extensive RC data and useful comments.
Thanks also to Joshua Adams and Erwin de Blok for providing relevant RCs and other files upon request, as well as Kristine Spekkens for fruitful discussions. 
The research was performed using the {\scriptsize PYNBODY} package \citep{Pontzen2013}, S{\scriptsize CI}P{\scriptsize Y} \citep{scipy}, N{\scriptsize UM}P{\scriptsize Y} \citep{numpy} and used {\scriptsize MATPLOTLIB} \citep{matplotlib} for all graphical representation.
The authors gratefully acknowledge the Gauss Centre for Supercomputing e.V. (www.gauss-centre.eu) for funding this project by providing computing time on the GCS Supercomputer SuperMUC at Leibniz Supercomputing Centre (www.lrz.de).
Some of this research was carried out on the High Performance Computing resources at New York University Abu Dhabi. 
We greatly appreciate the contributions of all these computing allocations.

\section*{Data Availability}
 
The data underlying this article will be shared upon request to the corresponding author(s).


\bibliographystyle{mnras}
\bibliography{example} 




\appendix

\section{The Diversity in the Stellar Mass-Velocity Diagram}\label{sec:MVDiagram}
\defcitealias{madore1987}{MW87}

We have thus far characterized and quantified the diversity in RCs amongst spiral galaxies along a variety of fundamental scaling relations. 
However, one wonders if a similar diversity is also matched in the light distribution.

In an effort to better grasp the stark relationship between baryons and DM in the TFR, \cite{madore1987}, hereafter \citetalias{madore1987}, explored the velocity-luminosity (V-L) plane, connecting spatially-resolved RCs with corresponding light profiles for 46 galaxies, and attempting to understand the projection of these profiles into the V-L space.
Any increase in the initial velocity profile should correspond to a matching increase in the light profile, producing a ``spatially resolved'' TFR.
\citetalias{madore1987}s analysis revealed hints of a ``universal'' slope within the inner regions of  galaxies. 
However, as we shall see below, our larger sample does not support such a conclusion.

In a similar vein, we study the M$^\star$-V$_{\text{circ}}$ plane, using stellar mass (M$^\star$) profiles for the PROBES and PROBES-II galaxies. 
Comparisons will also be made with NIHAO galaxies.   
The use of a stellar mass, as opposed to observed galaxy brightnesses in \magss, facilitates comparisons with galaxy evolution simulations, at the risk of additional systematic errors. 

\begin{figure}
    \includegraphics[width=\columnwidth]{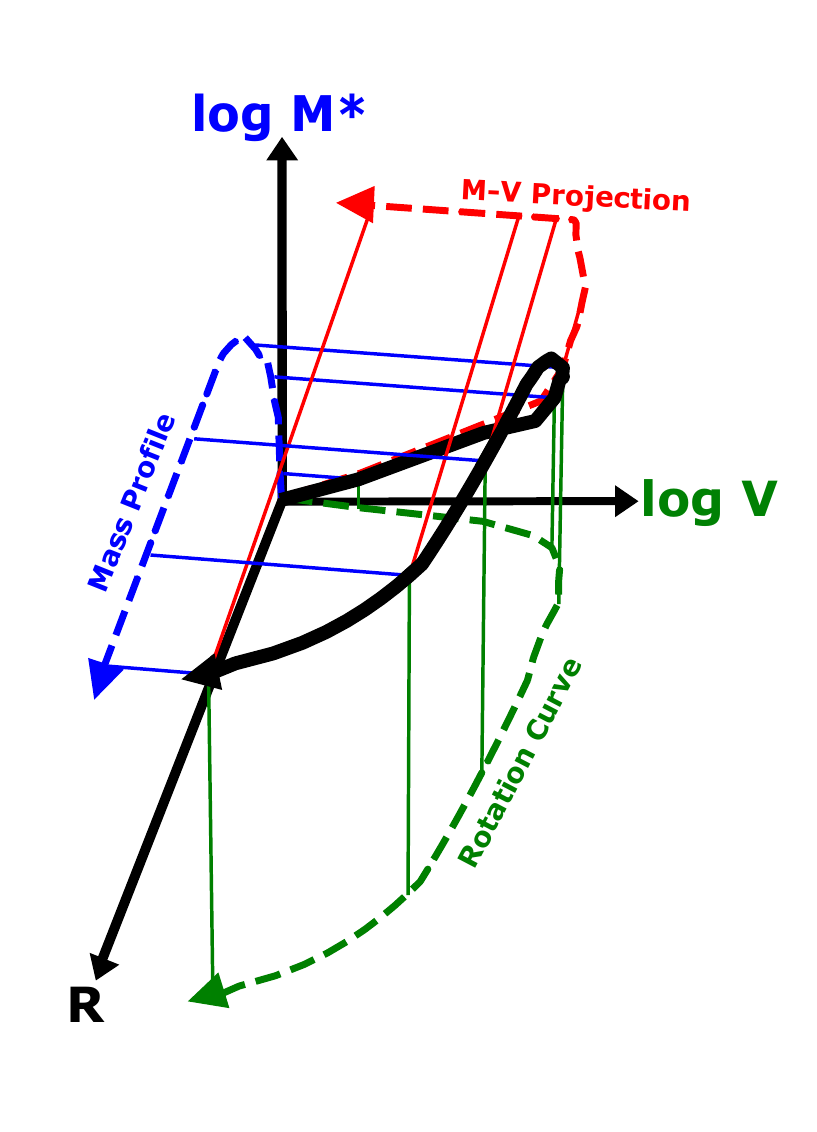}
    \caption{Adapted from MW87, the M$^\star$-V$_{\text{circ}}$-R space is presented for an idealised galaxy with a declining RC. 
    The dashed blue and green lines show the integrated mass profile and the spatially-resolved RC, respectively. 
    The red dashed line shows the corresponding projection into the stellar mass - rotational velocity (M$^\star$-V$_{\text{circ}}$) plane. 
    The solid black line shows the projection of these profiles into 3D space.}
	\label{fig:VLplot}
\end{figure}

\begin{figure*}
    \includegraphics[width=\textwidth]{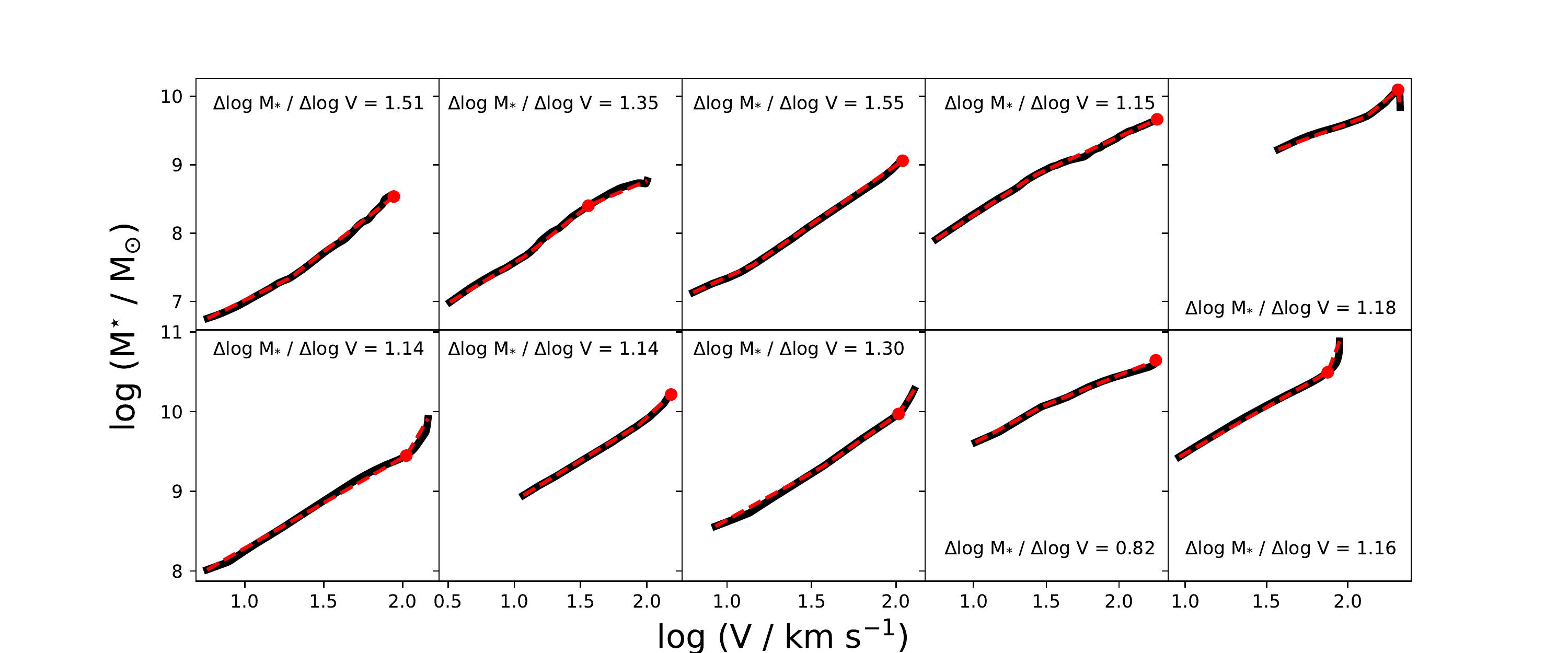}
    \caption{Projection of RC and mass profiles onto the M$^\star$-V$_{\text{circ}}$ space for 10 representative PROBES-II galaxies over a range of stellar masses. 
    The y-axis displays the enclosed stellar mass at a given radius, while the x-axis shows the rotational velocity of the galaxy at the corresponding radius. 
    A general consistency is seen in the Initial Linear Branch (ILB) within the inner regions of most galaxies. 
    The red point denotes the turning point of the galaxy off the ILB, while the red dashed line below is the RDP solution used to determine the turning point.}
    \label{fig:VMall}
\end{figure*}

The general morphology of the M$^{\star}$-V$_{\text{circ}}$ projection for galaxies is seen in \Fig{VLplot}. 
The RC and stellar mass profiles are represented by green and blue dashed lines, respectively, while
the red dashed and solid black lines show the resulting projection onto the M$^{\star}$-V$_{\text{circ}}$ plane and resulting profile in the 3D M$^{\star}$-V$_{\text{circ}}$-R space, respectively. 
The rotational velocity and the total enclosed stellar mass both initially increase linearly with radius. 
This region is referred to as the initial linear branch (ILB; \citetalias{madore1987}). 
As the RC transitions from solid-body rotation (${\rho} \propto r^{-1}$) to differential rotation (${\rho} \propto r^{-2}$), the RC flattens. 
Simultaneously, the growth of the total enclosed mass is reduced, as 
the optical surface brightness falls off roughly exponentially. 
Here, the M$^\star$-V$_{\text{circ}}$ projection curves up, becoming nearly vertical as the RC is constant, while the baryonic mass profile increases slightly.
This defining radius marks the end of the ILB, beyond which the RC and mass profiles no longer grow concurrently.
Finally, if the RC extends far enough (e.g., with long integration times), a decline in rotational velocity should be seen (${\rho}_{DM} \propto r^{-3}$). 
At that point, the M$^\star$-V$_{\text{circ}}$ projection curls back with a horizontal trajectory at constant integrated stellar mass towards the stellar mass axis. 
Examples of observed projections are seen in \Fig{VMall} for a random subset of ten galaxies in the PROBES-II sample. 

Our identification of the ILB of galaxies goes as follows. 
First, we sample the rotational velocity of the galaxy with 1000 points along the fitted MP model. 
Then, we linearly interpolate the necessary points in the stellar mass profile that correspond to the velocity at the same radii. 
The projections are normalized to lie within (-1, 1) and (-1, 1).
We then apply the Ramer–Douglas–Peucker (RDP) algorithm to the L-V and M$^\star$-V$_{\text{circ}}$ projections. 
This algorithm decimates a curve composed of line segments into a related curve consisting of fewer line segments. 
This is accomplished with a recursive function that identifies points that are further than a criterion $\epsilon$ from the segment obtained by any two points along the profile to be kept; 
points that are within the criterion can be safely discarded, as they lie close to the line segmenting the two points and are therefore redundant. 
In this study, we choose $\epsilon = 0.05$. 
This choice of $\epsilon$ tends to decimate the M$^\star$-V$_{\text{circ}}$ projection into 3 line segments, but provides leeway for projections that are better matched by 2 or 4 line segments. 
The smaller the $\epsilon$, the closer the final result to the original curve. 
Once the RDP algorithm is applied to the projection, the turning point at the end of the ILB is identified as the steepest angle between the simplified line segments. 
If no turning points are identified, the entire projection is considered to make up the ILB. 
The ILB slope is then calculated between the point with lowest velocity measurement and the identified turning point. 

We note that some PROBES-II galaxies do not exhibit purely linear relation between those two points, occasionally displaying a quick increase in rotational velocity from the galactic center before the mass profile starts increasing. 
This is also true with the larger PROBES dataset, but not for NIHAO galaxies which display purely linear ILB slopes. 
This non-linearity may be due to the linear interpolation of the mass profile, which is not necessary for NIHAO. 
The effect is sufficiently small that we can keep treating the ILB as linear. 
Ultimately, the ILB slope is a measure of the combined slopes of RC and (enclosed) light profile: a large ILB value describes a light profile that grows faster than the RC, while a small ILB value suggests RC growth at a rate faster than that of the light profile curve of growth.  

\citetalias{madore1987} found the slope of the ILB in the L-V plane to be remarkably consistent, with to $\Delta$M$_r$ / $\Delta$log V $= -3.30$ mag km s$^{-1}$, with a scatter of 0.41, where M$_r$ is the \textit{r}-band absolute magnitude, for most of their 46 galaxies.
However, PROBES-II and PROBES galaxies with RCs and light profiles give a different result with a mean ILB slope of $\Delta$M$_r$ / $\Delta$ log V $= -3.6$ mag km s$^{-1}$ with a scatter of 2.0. 
While our result encompasses that of \citetalias{madore1987}, its larger error is explained by our larger sample and greater range of stellar masses in PROBES-II. 
Hereafter, the L-V space is no longer considered, in favour of the readily testable M$^\star$-V$_{\text{circ}}$ space.


Examples of the M$^\star$-V$_{\text{circ}}$ distributions for 10 PROBES-II galaxies is shown in \Fig{VMall}.
For pure baryons in a self-gravitating spherical system, 
the theoretical expectation for the logslope of the M$^\star$-V$_{\text{circ}}$ projection is $\Delta$log M$^{\star}$ / $\Delta$log V $ = 2$. 
As most galaxies in this sample are DM dominated, any galaxy with DM should have a $\frac{d\log(M^\star)}{d\log(V)}$ slope below 2; indeed, PROBES-II galaxies have on average $\Delta$log M$^{\star}$ / $\Delta$log V $= 1.50$  with a scatter of $0.64$. 

We can also compute the M$^\star$-V$_{\text{circ}}$ slope for a pure exponential disk. 
The RC of an exponential disk is \citep{BandT1987}:
\begin{equation}
    V_c^2(R) = R\frac{\partial \Phi}{\partial R} = 4\pi G\Sigma_0R_d y^2\left[I_0(y)K_0(y) - I_1(y)K_1(y)\right],
\end{equation}
with a mass distribution of 
\begin{eqnarray}
    M(R) &=& 2\pi\int^{R}_0 \Sigma_0 \exp(-R^\prime/R_d)R^\prime dR^\prime\\
         &=& 2\pi \Sigma_0 R_d^2\left[1-\exp(-R/R_d)\left(1+\frac{R}{R_d}\right)\right]
\end{eqnarray}
where $R_d$ is the disk scale length, $y = R/2R_d$, and $I_n$ and $K_n$ are spherical Bessel functions of the $n$th kind. 
With these equations, we can numerically calculate the expected ILB slope for a pure baryonic exponential disk to be $\Delta \log$ M$^{\star}$ / $\Delta \log$ V $ = 2.35$. 
Again, these predictions are upper limits for pure baryon cases only; additional DM should lower this ratio.
The shape of the potential due to the baryons clearly influences the ILB slope. 

\begin{figure*}
	\includegraphics[width=\textwidth]{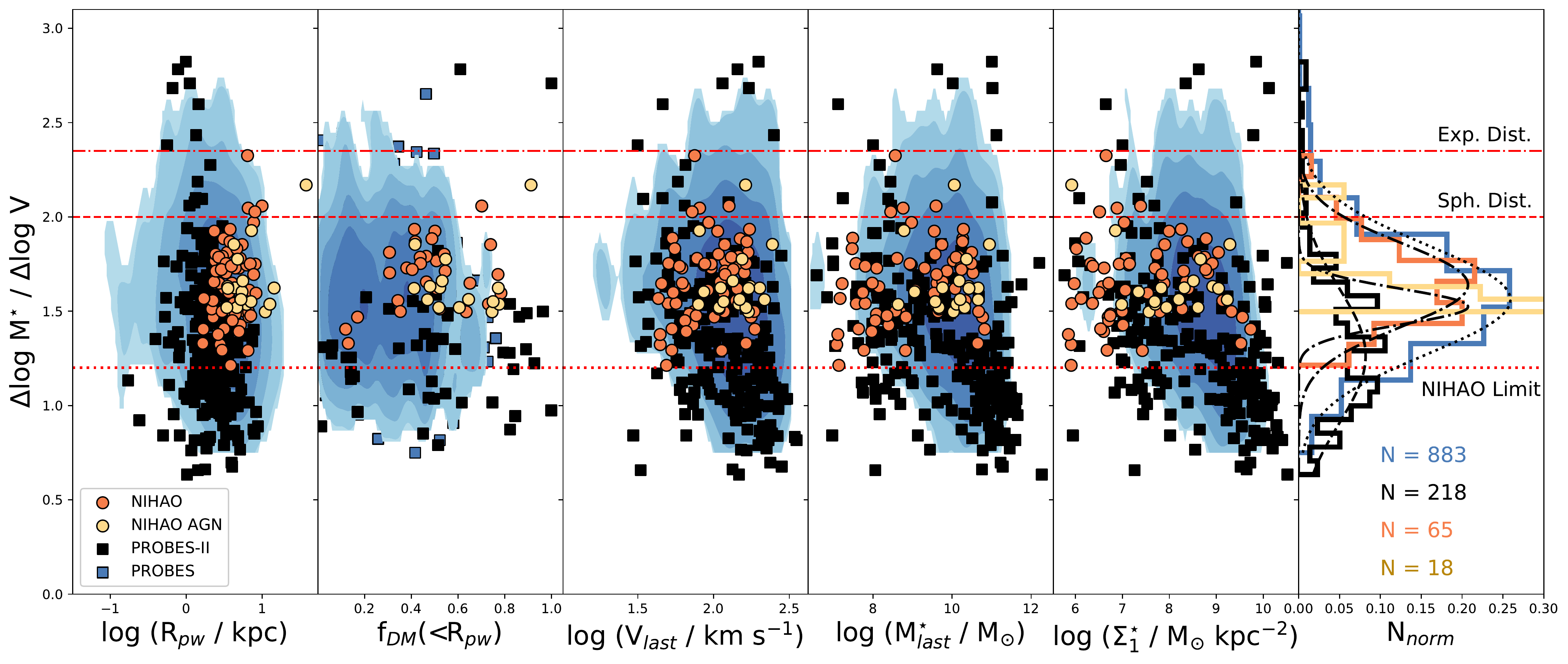}
    \caption{ILB slope of PROBES-II, PROBES, NIHAO, and NIHAO-AGN galaxies plotted with colours as in \Fig{NIHAOslopes}, against structural parameters as previously presented in \Fig{NIHAOslopes}. 
    The far right shows a normalized histogram of the samples fitted with Gaussian distributions; the dotted line following the PROBES distribution, the dashed line following the PROBES-II distribution, and the dash dotted lines following the NIHAO distributions. 
    The top red dashed dotted line indicates the ILB slope of 2.35 for a pure baryon exponential disk distribution, while the middle dashed line indicates the ILB slope of 2 for a spherical distribution of baryons. 
    The bottom dotted line indicates the lowest ILB slope for NIHAO galaxies, at 
    $\Delta$log M$^{\star} / \Delta$log V $= 1.2$.} 
    \label{fig:VMR_all}
\end{figure*}

The ILB slopes from NIHAO, NIHAO-AGN, PROBES, and PROBES-II are compared in \Fig{VMR_all} against the piecewise fit transition radius, R$_{pw}$, the average fraction of dark matter within R$_{pw}$, $f_{\text{DM}} (<R_{pw})$, the maximum rotational velocity, the last detected stellar mass value, last detected radius value (from the RC profile), and surface mass density at 1 kpc. 
The NIHAO and NIHAO-AGN galaxies follow the same general trends outlined by PROBES and PROBES-II galaxies. 
However, while NIHAO galaxies generally match the observed distribution of high ILB slopes, NIHAO galaxies with ILB slopes $\Delta$log M$^{\star}$ / $\Delta$log V $<  1.2$ are not found, possibly due to selection effects. 
Similarly, NIHAO-AGN galaxies are missing below $\Delta$log M$^{\star}$ / $\Delta$log V $\sim  1.5$, and this is partly due to the small number of simulated objects in that dataset.

We can further investigate the M$_\star$-V$_{\text{circ}}$ slopes predicted by NIHAO and NIHAO-AGN. 
The main difference between the simulated and observed projections is that the latter includes more slow rotators than the former. 
The distribution of ILB slopes for NIHAO above $\Delta$log M$^{\star}$ / $\Delta$log V $\sim  1.2$ matches the distribution of observed slopes quite well, as seen in the histograms in \Fig{VMR_all}. 
The NIHAO-AGN sample of 18 galaxies peaks at the same ILB slope value as the other datasets. 
However, both NIHAO simulations struggle to reproduce the low ILB end of the observed distributions. 
The low ILB regime contains galaxies whose RC slopes are mild relative to that of their enclosed stellar mass profile. 
That neither NIHAO nor NIHAO-AGN match observations below $\Delta$log M$^{\star}$ / $\Delta$log V $\sim  1.2$ may result from the small number of simulated systems;  larger samples will help in establishing how much of these discrepancies are due to statistics or genuine shortcomings of the physics of galaxy formation.

Interestingly, no NIHAO galaxies are found above the theoretical prediction for pure exponential baryonic disks, despite a small number of observed galaxies in that range.
Observed galaxies above this upper limit have therefore been scattered high by observational errors \citep{2021ApJ...912...41S}. 

Even in cases where systematic observational errors can be ignored (e.g., NIHAO simulations), there is greater scatter in $\Delta$log M$^{\star}$ at fixed $\Delta$log V, than there is scatter in $\Delta$log V at fixed $\Delta$log M$^{\star}$. 
This is also observed in both observed data sets. 
This again suggests that the M$^\star$-V$_{\text{circ}}$ projection is mainly driven by variations in the stellar mass profile.
There is typically twice as much scatter in $\Delta$log M$^{\star}$ than there is in $\Delta$log V.

Observational errors in \Fig{VMR_all} were estimated using the technique of \cite{2021ApJ...912...41S}. 
Gauging by the excess spread of the observed data over the NIHAO and NIHAO-AGN distributions, 
systematic errors for both PROBES data sets are comparable. 
1$\sigma$ errors in distance, inclination, observed rotation velocity, surface brightness, and stellar mass-to-light ratio are propagated through the analysis of all PROBES and PROBES-II galaxies, and the average resulting displacements along the $\Delta$log V versus $\Delta$log M$^{\star}$ relation are rather minimal.
This suggests that observational errors alone cannot account for the discrepancy in scatter between the observed and simulated galaxies.
The simulated galaxies do not reproduce the full range of scatter seen in observed galaxies, and they thus do not reproduce the observational diversity in these parameter spaces. 


\bsp	
\label{lastpage}
\end{document}